\documentstyle[preprint,aps,tighten,epsf]{revtex}

\begin{document}

\thispagestyle{empty}

\title{The $p p \rightarrow p p 
\pi^0$ Reaction near  Threshold: A Chiral Power Counting Approach}

\author{Thomas D. Cohen$^a$, 
James L. Friar$^b$, 
Gerald A. Miller$^c$, and
Ubirajara van Kolck$^c$ \vspace{15pt} }
\address{$^a$ Department of Physics\\ University of Maryland\\ 
College Park, MD 20742 \vspace{15pt} } 
\address{$^b$ Theoretical Division\\ Los Alamos National 
Laboratory\\ Los Alamos, NM 87545 \vspace{15pt} }
\address{$^c$ Department of Physics\\ University of 
Washington, Box 351560\\ Seattle, WA 98195-1560 \vspace{15pt}}

\maketitle

\begin{abstract} 
 We use  power-counting arguments as an organizing principle
to apply chiral perturbation theory, including an explicit 
$\Delta$, to the  $p p \rightarrow p p \pi^0$ reaction near 
threshold. There are two lowest-order leading mechanisms 
expected to contribute to the  amplitude with similar magnitudes: 
an impulse term, and a $\Delta$-excitation mechanism.  We examine 
formally sub-leading but potentially large mechanisms, including 
pion-rescattering and short-ranged contributions.  We show that the
pion-rescattering contribution is enhanced by off-shell effects and  
has a sign opposite to that of a recent estimate based on a PCAC 
pion interpolating field. Our result is that the impulse term 
interferes destructively with the pion rescattering and 
$\Delta$-excitation terms. In addition, we have modeled the 
short-ranged interaction using $\sigma$ and $\omega$ exchange mechanisms. 
A recoil correction to the impulse approximation is small.
The total amplitude obtained including all of these processes is 
found to yield cross sections substantially smaller than the measured 
ones.
\end{abstract}

\vspace{2cm}

\hfill{DOE/ER/40427-26-N95}

\hfill{DOE/ER/40762-074, U of MD pp 96-058}
\newpage

\section{Introduction} 
The theory of the threshold behavior of the reaction  
$p p \rightarrow p p \pi^0$ has been studied for some time
\cite{KR66,SSY69,MS91,N92,LR92,HGM94,HO95,julich}.
The numerical results of the early analyses \cite{KR66,SSY69,MS91,N92} 
were that this behavior is dominated by the impulse term (Fig.(\ref{f1}a)).
In this process, a single pion is emitted {}from a nucleon. Effects of 
initial- and final-state interactions are taken into account by using 
scattering wave functions that solve the Schr\"odinger equation 
incorporating the full nucleon-nucleon potential. 
In such a process at threshold, the pion will emerge in an s-wave.  
In the early calculations the main competitor process was thought to be  
``pion rescattering'' {}from a pion seagull term in which the nucleon emits 
two pions and one of them is re-absorbed by the other nucleon, 
as illustrated in Fig.(\ref{f1}b). The early estimate of this 
pion-rescattering term gave a small contribution \cite{KR66}
and it was believed that the impulse term alone could account for 
the essential features of the data.

Good data for the $p p \rightarrow p p \pi^0$ reaction near 
threshold became available far later than these original 
theoretical estimates \cite{M90,uppsala}. These data clearly show that 
the early theoretical calculations were inadequate.  While the impulse 
term can account for the energy dependence of the cross section 
(once the phase space and Coulomb effects are treated 
properly) \cite{MS91,N92}, the total cross-section is about a 
factor of five below the data.  Clearly something is missing in 
the original theoretical analysis. To the best of our knowledge there 
have been two proposed explanations for the missing physics.

The first explanation, due to Lee and Riska \cite{LR92}, is that 
the original analysis is lacking important effects due to exchanges 
of shorter range.  In particular, they suggest that an effect due to 
$\sigma$ and $\omega$ meson exchanges such as that shown in Fig.(\ref{f1}c)  
can account for the discrepancy.  They find that if 
they use meson-nucleon coupling constants and meson masses obtained {}from 
meson-exchange potentials this process plus the impulse approximation
reproduces the data within good accuracy.  
Their analysis was confirmed in the work of Horowitz, 
Griegel and Meyer \cite{HGM94}.  This explanation is interesting 
in several ways: it is perhaps the first example of a shorter 
than pion-ranged meson-exchange effect playing such a critical 
role in an observable and it is based on a ``z-graph'' effect of 
scattering into and out of a negative energy state. Hence, the latter
could provide further evidence for the validity of the Dirac phenomenology 
used in intermediate-energy $p$-nucleus collisions \cite{W87}.

A second explanation has recently been proposed by Hern\'andez and 
Oset \cite{HO95}.  They observed that the 
early estimates of the pion-rescattering term were based on the 
on-shell $\pi N$ scattering amplitudes, whereas in the 
contribution to the $p p \rightarrow p p \pi^0$ reaction one of 
the pions is well off-shell.  Using models based on a pion 
interpolating field satisfying the partially conserved axial 
current (PCAC) requirement, they estimate that the effect of the 
off-shell behavior  of the amplitude in the pion-rescattering 
term, together with the impulse term, is large enough to account for 
the data,  although they note that at present the uncertainties 
are large.

Which of these explanations, if either, is the correct one?  Clearly 
both effects cannot simultaneously be as large as the authors of the 
papers suggest unless there are some compensating effects. For example, 
if the amplitude of the impulse term were added to the amplitude 
for both of these effects, the resulting cross section would be 
approximately 2--2.5 times larger than the experimental cross section. 
A fundamental difficulty in trying to assess whether either of 
these explanations is correct is that almost all of the work on the 
subject to date has been somewhat {\it ad hoc} in an important sense.  
The analyses have not been based on a systematic expansion for which one
can account for all processes without double counting, and for which 
one has some {\it a priori} method of deciding which processes should 
be small.  In short, there has been no simple organizing principle. 

This lack of an organizing principle has led to a reliance on intuition 
in deciding which processes are likely to be important. There are an 
infinite number of processes one can consider and the question is which 
processes will make significant contributions.   Unfortunately, intuition 
can be faulty.  For example, consider the process in Fig.(\ref{f1}d) 
in which the interaction generates a $\Delta$ that
emits a pion and becomes a nucleon. This process has not been 
included in any of the analyses to date. There is a simple 
intuitive reason why such an effect should be small \cite{M90}:  
the $\pi N \Delta$ vertex proceeds only through p-waves 
in the center of mass of the $\Delta$.  At threshold, the pion is 
in an s-wave relative to the center of mass of the entire system. 
Thus, one can get a nonvanishing amplitude for this process only 
because the center of mass of the $\Delta$ may differ 
{}from the center of mass of the system, and one expects such 
differences to be small\cite{N92}.  Indeed, we estimate this effect below
and find it  small  compared to typical hadronic amplitudes.  
However, {\it all} of the effects discussed so far are small in this  
sense.  When we evaluate this  effect numerically its amplitude is 
found to be sensitive to the choice of the potential used to produce 
the scattering wave functions, being in some cases similar in magnitude 
to these other effects.
 
The fact that all of these effects are small is no coincidence.  
Indeed, it is a straightforward consequence of approximate chiral symmetry.  
Suppose, hypothetically, that we lived in a chiral world in which the 
up- and down-quark masses were exactly zero. It is easy to 
establish the theorem that the amplitude for emission of a pion at 
threshold must vanish.  Of course, in the real world the pion mass is 
not zero---but it also  is not large compared to typical hadronic 
scales.  This suggests that the amplitude can be described in a 
systematic manner as an expansion in ($Q/M$), where
$Q$ is a typical (small) momentum or energy scale
and $M$ is a typical (large-mass) hadronic scale $\sim$ 1 GeV, 
such as $m_N$, $m_\rho$ or 4$\pi f_\pi$. This systematic description is 
called chiral perturbation theory ($\chi$PT).

For the example just described, $Q \sim m_{\pi}$ characterizes the behavior 
of the outgoing pion and one has $m_\pi / M \sim$ 0.1 -- 0.2 .
Our problem has an additional scale that is less obvious (although
certainly well known). The internal nuclear momentum is quite large, since
the pion mass must be generated entirely {}from the nuclear kinetic energy. 
We will show below that this corresponds to $Q \sim (m_\pi M)^{1/2}$. 
This results in a larger than desirable expansion parameter, 
$(m_\pi/ M)^{1/2} \sim$ 0.4, but one that nevertheless will allow a 
systematic expansion.

The simplest way to implement $\chi$PT is via effective 
Lagrangians aided by power-counting arguments.  The seminal idea 
was contained in a paper by Weinberg \cite{W79}.  This idea was 
developed systematically for interactions of mesons \cite{GL84} 
and for interactions of mesons with a baryon \cite{JM91,BKM95}.  
The generalization of these techniques to 
describe properties of more than one  baryon was also due to 
Weinberg \cite{W90} and was carried out in detail in Refs.\cite{ORV94,V94}.  
Here we will use Weinberg's power-counting arguments to 
organize a calculation of the threshold 
production of a $\pi^0$ in a $p p$ collision.

 The general idea of power counting in effective Lagrangians is, 
in fact,  far more powerful than $\chi$PT.  In principle, one can 
include in the Lagrangian and the power counting any light degree of 
freedom (and not simply the Goldstone excitations). 
Indeed, the failure to treat all relevant degrees of freedom explicitly 
may well lead to very slow convergence 
for the expansion and may limit its usefulness.  In the case of baryons
an obvious light degree of freedom is the $\Delta$, since 
$m_\Delta-m_N$ is small, being about 2 $m_\pi$, and the $\Delta$ 
is strongly coupled to the $\pi N$ channel. 
There remains some controversy 
over whether or not one should include the $\Delta$ explicitly.  
We strongly advocate the position that the $\Delta$ should be included 
explicitly.  Moreover, we will show by explicit calculation that the 
leading-order amplitude associated with an explicit $\Delta$ is
similar in magnitude to the impulse approximation and
makes significant contributions.
Hence, in our counting we will consider  $m_\Delta - m_N$ to be of 
order $m_\pi$. We note that alternative approaches assume 
that this quantity is of order $M$. For pion production reactions
a virtual $\Delta$ can have a small energy denominator ($\sim m_\pi$), 
and it makes little practical sense to treat that particle 
as far off shell.

The power counting has considerable power in organizing the calculation.
It tell us which processes are expected to dominate.
But it also provides strong constraints on the form of the low-energy
Lagrangian, which in turn is reflected in constraints on effects of the
off-shell dependence of the various sub-processes, in particular
the seagull or pion rescattering.
In principle, one ought to be able to apply the power counting in 
a completely systematic
manner and work consistently to some given order.  The effective 
theory has a finite number of free parameters at a given order 
that one fits {}from some set of observables.  Once determined one 
can predict other observables.  This approach has been used
with great success in describing the properties of the 
pseudoscalar mesons \cite{GL84}.  In the present case,
for technical reasons which we will discuss later, this procedure 
cannot be implemented in a practical way.  Instead,
we will use power counting as a guide to which processes we 
expect to be large.

This article is organized as follows:  
In the following section we briefly review Weinberg's power counting 
scheme in general.  
Next, we discuss special features of this scheme for the present 
problem. Various  terms are considered.
We then derive the operators associated with the various leading-order  
processes in $\chi$PT: the impulse, $\Delta$-excitation, pion-rescattering, 
recoil correction, and short-range mechanism as modeled by $\sigma$
and $\omega$ exchanges. 
In the same section these amplitudes are evaluated using
three different potentials to describe the initial and final state 
$pp$ interactions. 
The impulse term interferes destructively with the $\Delta$-excitation, 
and pion-rescattering amplitudes. The resulting cross sections 
that we compute are found to be far smaller than the measured 
ones, even when the $\sigma, \omega$-exchange term  is included. Next 
these results are discussed, and the difference between the 
off-shell behavior of our seagull term and that seen in Ref.\cite{HO95}
is explained. The paper ends with a discussion of a possible direction 
for future research using $\chi$PT  for pion production reactions.

A preliminary report of these results was presented elsewhere \cite{BLO95}.
Before this manuscript was completed, a paper appeared \cite{SC95} 
that also uses $\chi PT$ to reach the same conclusion
regarding the interference between impulse and seagull terms.
Our work differs {}from theirs in that we modify Weinberg's power
counting to the particular kinematics of this problem, and that we
consider the $\Delta$ explicitly.

\section{Review of Weinberg's power counting}

Three-momenta $Q$ exchanged in {\it typical} nuclear systems are on the
order of the pion mass, $Q \sim m_{\pi}$, which is small compared to
the characteristic QCD mass scale: $M \sim$ 1 GeV. Whenever we face such a
two-scale problem it is useful to separate the corresponding
physics by considering an effective, low-energy theory that involves
only the relevant degrees of freedom, all with small three-momenta
$Q$. The pion and the nucleon obviously play this role,
since they are the lightest stable (with respect to strong interactions)
hadrons. Low-lying resonances also ought to be relevant. Prominent
among them is the $\Delta$ isobar of mass $m_{\Delta}= m_{N}+\delta$,
with $\delta \simeq 2m_{\pi}$, which is comparable to $Q$ \cite{JM91}.  
Moreover, consistency of the chiral expansion with a large-$N_c$ 
expansion requires an explicit $\Delta$ \cite{CB92,C95a,C95b}. 
The case for higher-mass baryon states is less clear, since 
their mass differences with respect to the nucleon are larger, 
and they couple more weakly to nucleons and pions. 
Higher-mass meson states also have masses comparable to $M$.
In the following we will, therefore, take the view that the degrees
of freedom that must be accounted for explicitly are the pion, the
nucleon and the isobar, while effects of higher-mass states are
included indirectly, as contributions to the several parameters of
the effective theory. 

The low-energy parameters are not necessarily small, so the only potential
expansion parameter is $Q/M$. ($Q$ stands not only for typical three-momenta,
but also for factors of $m_{\pi}$ and $\delta$.) 
Note that, because we are restricted to small three-momenta, nucleons
and isobars are non-relativistic and act very much like static sources
of pions. Corrections to the static
limit can be accounted for by an expansion in $Q/m_{N}$.

The next task is to count powers of $Q$ in an arbitrary diagram. 
This is simple to do in the case of diagrams with only pions \cite{W79}, 
and can be straightforwardly extended to diagrams with one nucleon. 
However, systems with several nucleons require more care, due to the 
appearance of infrared enhancements in reducible graphs \cite{W90}.
Diagrams where all energy denominators are of order $Q$ are
called irreducible. In this case the same power counting applies:
an irreducible diagram with $V_{i}$ vertices of type $i$, $L$ loops,
$C$ separately connected pieces, and $E_{f}=2A$ external fermion lines, 
will be proportional to $Q^{\nu}$, with
\begin{equation}
 \nu=4-A+2L-2C+\sum_{i=1}^{\infty} V_{i}\Delta_{i}.      \label{pc1}
\end{equation}
\noindent
Here the so-called index of a type $i$ vertex is defined as 
\begin{equation}
 \Delta_{i}=d_{i}+\frac{f_{i}}{2}-2,             \label{pc2}
\end{equation}
in terms  of $d_i$ (which is the sum of the number of derivatives,
the number of powers of $m_{\pi}$, 
and the number of powers of $\delta$)
and of $f_{i}$ (which is the number of fermion field operators).

However, because of the non-relativistic character of nucleons, there also
exist graphs with intermediate states that differ in energy {}from
initial or final states by only a small amount of order
$Q^{2}/2m_{N}$. They are larger than irreducible diagrams by factors 
of $m_{N}/Q$. We call these diagrams reducible; they can be
split into irreducible sub-diagrams by cutting only lines corresponding
to initial or final particles.  
The sum of irreducible diagrams is just what is usually called the potential. 
Reducible diagrams can be obtained
by iteration of the potential, generating wave functions corresponding
to scattering or bound states. In processes such as the ones in which we are
interested, the amplitude can be organized into
an irreducible part (to which the external pions are attached)
sandwiched between wave functions of the initial and final nuclear
states, which now contain all of the reducible parts.

Now, if $\Delta_{i} \geq 0$ for all $i$, 
then a perturbative expansion in $Q/M$
exists for irreducible diagrams. Chiral symmetry provides exactly
this constraint. All evidence suggests that QCD has an
approximate $SU(2)\times SU(2)\sim SO(4)$ symmetry that is spontaneously 
broken to $SU(2)\sim SO(3)$. The pions are the Goldstone bosons
associated with this breaking; as such, there is at least one choice
of fields for which in the chiral limit ($m_{\pi} \rightarrow 0$)
they couple via derivatives to other particles and themselves
\cite{CWZ}, and this is sufficient to guarantee $\Delta_{i} \geq 0$.

The interactions of the effective low-energy Lagrangian can thus be 
ordered according to the index in Eq.(\ref{pc2}). 
Below we present only those terms that
are directly relevant to our subsequent calculation; in particular we
subsume in ``$\cdots$'' those interactions with additional pion fields 
that are necessary to construct the correct (nonlinear) realizations of 
chiral symmetry, but are not required in the lower-order calculations 
that we will perform. 
The lowest-order Lagrangian is the one where $\Delta_{i} = 0$ 
for each interaction \cite{W79,GL84,JM91,BKM95,W90,ORV94,V94},
\newcommand{\boldpi}{\mbox{\boldmath $\pi$}}
\newcommand{\boldtau}{\mbox{\boldmath $\tau$}}
\newcommand{\boldT}{\mbox{\boldmath $T$}}
\begin{eqnarray}
 {\cal L}^{(0)} & = & 
          \frac{1}{2}(\dot{\boldpi}^{2}-(\vec{\nabla}\boldpi)^{2})
          -\frac{1}{2}m_{\pi}^{2}\boldpi^{2} \nonumber   \\
    &   & +N^{\dagger}[i\partial_{0}-\frac{1}{4 f_{\pi}^{2}} \boldtau \cdot
         (\boldpi\times\dot{\boldpi})]N +\frac{g_{A}}{2 f_{\pi}} 
         N^{\dagger}(\boldtau\cdot\vec{\sigma}\cdot\vec{\nabla}\boldpi)N
                                               \nonumber \\
    &   & +\Delta^{\dagger}[i\partial_{0}- \delta]\Delta 
          +\frac{h_{A}}{2 f_{\pi}}[N^{\dagger}(\boldT\cdot
          \vec{S}\cdot\vec{\nabla}\boldpi)\Delta +h.c.] +\cdots \, , 
\label{la0}
\end{eqnarray} 
\noindent
where $f_{\pi}=93$ MeV is the pion decay constant,
$\delta = m_{\Delta}-m_{N}$ is the isobar-nucleon mass difference,
$g_{A}$ is the axial-vector coupling of the nucleon, $h_{A}$
is the~$\Delta N \pi$ coupling, and $\vec{S}$ and $\boldT$ are the 
transition spin and isospin matrices, normalized such that
\begin{eqnarray}
  S_{i}S^{+}_{j} & = & \frac{1}{3} (2\delta_{ij} - 
              i\varepsilon_{ijk} \sigma_{k})    \label{S1}   \\
  T_{a}T^{+}_{b} & = & \frac{1}{3} (2\delta_{ab} 
              - i\varepsilon_{abc} \tau_{c}).   \label{S2}
\end{eqnarray}
\noindent
Notice that we defined the fields $N$ and $\Delta$ in
such a way that there is no factor of $exp(-i m_N t)$ in their
evolution. Hence $m_N$ does not appear explicitly to this order:
the baryons are static.  
We also wrote ${\cal L}^{(0)}$ in the rest frame
of the baryons, which is the natural choice. (Galilean invariance
will be assured by including terms with additional derivatives.) 
Chiral symmetry determines the coefficient of the so-called
Weinberg-Tomozawa term 
($N^{\dagger}\boldtau \cdot(\boldpi\times\dot{\boldpi})N$)
but not of the single-pion interactions ($g_A, h_A$).
              
The first-order Lagrangian has $\Delta_{i} =1$ 
\cite{JM91,BKM95,ORV94,V94},
\begin{eqnarray}
 {\cal L}&^{(1)}& 
        =\frac{1}{2m_{N}}[N^{\dagger}\vec{\nabla}^{2}N
        +\frac{1}{4 f_{\pi}^{2}}(iN^{\dagger}\boldtau\cdot
        (\boldpi\times\vec{\nabla}\boldpi)\cdot\vec{\nabla}N + h.c.)] 
                                       \nonumber \\
  &   & +\frac{1}{f_{\pi}^{2}}N^{\dagger}[(c_2 +c_3 - \frac{g_A ^2}{8 m_{N}})
        \dot{\boldpi}^{2} -c_3 (\vec{\nabla}\boldpi)^{2} 
        -2c_1 m_{\pi}^{2} \boldpi^{2} -
        \frac{1}{2} (c_4 + \frac{1}{4m_{N}}) 
        \varepsilon_{ijk} \varepsilon_{abc} \sigma_{k} \tau_{c} 
        \partial_{i}\pi_{a}\partial_{j}\pi_{b}]N \nonumber   \\  
  &   & +\frac{\delta m_{N}}{2} N^{\dagger}[\tau_{3}-\frac{1}{2 f_{\pi}^{2}}
        \pi_3 \boldpi\cdot\boldtau]N  
        +\frac{1}{2m_{N}}\Delta^{\dagger}[\vec{\nabla}^{2} +\cdots]\Delta 
                                                               \nonumber \\
  &   & -\frac{g_{A}}{4 m_{N} f_{\pi}}[iN^{\dagger}\boldtau\cdot\dot{\boldpi}
        \vec{\sigma}\cdot\vec{\nabla}N + h.c.]             
        -\frac{h_{A}}{
        2 m_{N} f_{\pi}}[
        iN^{\dagger}\boldT\cdot\dot{\boldpi}\vec{S}\cdot\vec{\nabla}
        \Delta + h.c.]                       \nonumber \\
  &   & -\frac{d_1}{f_{\pi}} 
        N^{\dagger}(\boldtau\cdot\vec{\sigma}\cdot\vec{\nabla}\boldpi)N\,
        N^{\dagger}N
        -\frac{d_2}{2 f_{\pi}} \varepsilon_{ijk} \varepsilon_{abc} 
        \partial_{i}\pi_{a}  
        N^{\dagger}\sigma_{j}\tau_{b}N\, N^{\dagger}\sigma_{k}\tau_{c}N 
        +\cdots \, ,             \label{la1}
\end{eqnarray}
\noindent
where the $c_{i}$'s are coefficients of ${\cal O}(1/M)$, 
$\delta m_{N} \sim m_d -m_u$ is the quark mass difference contribution to 
the neutron-proton mass difference,
and the $d_{i}$'s are coefficients of ${\cal O}(1/f_\pi^2 M)$. 
These seven numbers are not fixed by chiral symmetry, but it is important 
to point out that Galilean invariance requires that the other coefficients 
explicitly shown above be related to those appearing in ${\cal L}^{(0)}$. 
This in particular
fixes the strength of the single-pion interactions in terms of the 
lowest-order coefficients $g_A$ and $h_A$, and of the common mass $m_{N}$.

The second-order Lagrangian, with $\Delta_{i} =2$, is
\begin{eqnarray}
{\cal L}^{(2)} & = & \frac{d_1^{\prime}+e_1}{2 m_N f_{\pi}} 
                 [iN^{\dagger}\boldtau\cdot\dot{\boldpi}
                 \vec{\sigma}\cdot\vec{\nabla}N\, N^{\dagger}N + h.c.] 
                                                   \nonumber \\
           &  & -\frac{e_1}{2 m_N f_{\pi}}
                 [iN^{\dagger}\boldtau\cdot\dot{\boldpi}\vec{\sigma}N\,
                 \cdot N^{\dagger}\vec{\nabla}N + h.c.] \nonumber \\
           &  & +\frac{e_2}{2 m_N f_{\pi}}
                 [N^{\dagger}\boldtau\cdot\dot{\boldpi}\vec{\sigma}\times
                 \vec{\nabla}N\,
                 \cdot N^{\dagger}\vec{\sigma}N + h.c.] 
                 +\cdots \, ,                  \label{la2}
\end{eqnarray}
\noindent
where the $e_i$'s are other coefficients of ${\cal O}(1/f_\pi^2 M)$. 

\section{Power counting and the $p p \rightarrow p p \pi^0$ 
reaction}

The power-counting arguments can be extended to the  
$p p \rightarrow p p \pi^0$ reaction.
However, there is one fundamental difference relative to Weinberg's 
standard power
counting due to the kinematics of the present problem. In the 
standard power-counting arguments it is assumed that typical 
momenta carried by nucleons are $\sim m_\pi$.  However, this
condition cannot be satisfied for the $p p \rightarrow p p \pi^0$ 
reaction.  Instead one finds:
\begin{equation}
p_{\rm typ} \sim \sqrt{m_N m_\pi} \, .
\label{typ}
\end{equation}
The reason for this is quite simple.  Consider the initial state 
in the center of mass frame.  At threshold the total energy is 
$2 m_N + m_\pi$,  
so that the initial kinetic energy of each nucleon is 
$m_\pi/2$.  Since the energy is small the nonrelativistic kinetic 
energy formula should apply, $p_{\rm typ}^2/(2 m_N) = m_\pi/2$, and
Eq.(\ref{typ}) follows.

While having typical momenta of order $(m_N m_\pi)^{1/2}$
can alter the details of the power counting, it should not spoil 
the scheme.  Indeed, 
the scale of the typical momenta still goes to zero as we 
approach the chiral limit and hence a chiral expansion remains 
sensible.  On the other hand these momenta are characteristically larger 
than what one usually encounters 
while chiral power counting and thus the expansion can be expected to 
be more slowly convergent than a more typical case.
In particular the expansion parameter becomes $\sqrt{m_{\pi}/M}$ 
rather than $m_{\pi}/M$.
The point is that since a typical momentum of the 
nucleons is $(m_N m_\pi)^{1/2}$ nothing prevents a momentum 
transfer of order $(m_N m_\pi)^{1/2}$ in interactions. Consider 
for example processes involving a single-pion exchange.  Chiral 
symmetry requires that (to leading order), each vertex contains
a derivative and hence is proportional to the momentum transfer. 
In the traditional Weinberg  power counting a  pion-nucleon vertex  
contributes $(m_{\pi}/M)$ to the total power counting of an irreducible 
graph.  With our kinematics, however, those vertices would 
contribute $(m_{\pi} m_N)^{1/2} / M \sim (m_{\pi}/ M)^{1/2}$, 
where the second form follows {}from $m_N \sim M$. 
Similarly, in the traditional power counting a 
meson propagator $1/(q^2 - m_\pi^2)$ contributes as $1/m_\pi^{2}$ 
while with our kinematics it goes as $1/(M m_{\pi})$.  In 
short, with these kinematics whenever a three momentum enters 
into the power counting it contributes to the power counting as 
$(m_{\pi} M)^{1/2}$.

There is another subtlety in the present power-counting scheme 
and this concerns the notion of irreducibility.  It should be recalled 
that a sub-diagram is reducible in Weinberg's sense if it includes 
a small energy denominator $\sim m_\pi^2/m_N$ and is irreducible 
otherwise. This leads 
to the possibility that a sub-diagram may look topologically as 
though it were
reducible, in the sense that one could cut the sub-diagram into 
smaller sub-diagrams with only nucleon external legs, while in 
fact the diagram is irreducible.  This happens if the kinematics 
requires that the energy denominator associated with cutting the 
sub-diagram is order $m_\pi$ and is not $\sim m_\pi^2/m_N$.  This 
is precisely what happens for the impulse term.

There is finally a caveat. In principle, in order to implement the power
counting in the irreducible piece, one ought to use a potential obtained 
{}from the same chiral Lagrangian according to the power-counting rules.  
While such a potential has been developed to third order \cite{ORV94} 
it is only successful at energies well below the threshold for pion 
production.  Accordingly, we will follow the strategy advocated 
by Weinberg in his treatment of the three-body 
problem \cite{W90} and use semi-phenomenological potentials that 
incorporate  experimental information into the nucleon-nucleon 
interaction. This strategy has a conceptual cost: there may be a 
mismatch between the nucleon field used in the
$NN$ potential and the one based on the chiral Lagrangian used in 
our calculation of the operators.  To attempt to estimate the 
scale of the uncertainty due to this we will use a number of 
different $NN$  potentials.
These effects are not negligible.

Let us now detail how this modified power counting works
for the irreducible diagrams close to threshold, 
where we can restrict ourselves to s-waves. We examine each of the 
various contributions.

Consider first the impulse term, Fig.(\ref{f1}a).  
At first glance this appears to be order 
$m_{\pi}^{3/2}/(f_{\pi} M^{1/2})$. 
It has been recognized that the s-wave amplitude is small because      
in Eq.(\ref{la0}) a pion of momentum $q$ couples to baryons only via  
$\vec{\sigma}\cdot \vec{q}$. Close to threshold, 
the interaction consequently
proceeds via the Galilean term $\partial_0 \boldpi$ in Eq.(\ref{la1}).
This yields an explicit factor of $m_\pi/f_{\pi}$,  
but also involves $1/m_N$ times a gradient  that contributes 
$p_{\rm typ}$.  Thus, the net 
contribution at the vertex is $m_{\pi}^{3/2}/(f_{\pi} M^{1/2})$.  
However, this is not yet the correct order for this process  
because of the subtlety  associated with the
concept of irreducibility discussed above.
Since the outgoing pion 
carries an energy of the order of the pion mass,
the energies of the $NN$ intermediate state before and after pion 
emission differ by $\sim m_\pi$.  
Therefore both of the intermediate states cannot simultaneously be 
within $\sim m_\pi^2/m_N$ of being on-shell:
at least one intermediate state, before or after emission,
is off shell by $\sim m_\pi$.

This single, relatively-high-momentum ($\sim \sqrt{m_{\pi} m_N}$) 
pion exchange must therefore be included in the 
irreducible class of operators for our process (unlike the usual case).
All other initial- and final-state interactions will be considered 
reducible and included in the wave functions.
Thus the irreducible sub-diagrams of Fig.(\ref{f1}) should 
in lowest order be drawn as in Fig.(\ref{f2}). 
The two--nucleon 
interaction itself provides a factor $1/f_{\pi}^2$. Indeed, if it
originates {}from virtual (static) pion exchange (Fig.(\ref{f2}a,b)), 
it results {}from two factors of
$\sqrt{m_\pi m_N}/ f_{\pi}$ {}from each vertex and one of 
$(m_\pi m_N)^{-1}$ {}from the propagator; 
if it arises {}from exchange of a heavier meson $h$
(Fig.({\ref{f2}c,d)), it is of order $g_{NNh}^2/m_h^2$
which is typically 
$\sim (4\pi/1 \mbox{GeV})^2 \sim 1/f_{\pi}^2$.
The inclusion of the interaction in the irreducible part also
produces an energy denominator between the pion exchange and the 
pion emission.  This energy denominator is $(E_{\rm intermediate} - 
E_{\rm initial})^{-1} \sim  kinetic \;\, energy \sim m_\pi$.  
So, along with the explicit factor of order 
$m_{\pi}^{3/2}/(f_{\pi} M^{1/2})$ 
{}from the pion emission vertex, we have an energy denominator of 
order $1/m_\pi$ and a factor of $ 1/f_{\pi}^2$ {}from the potential. 
This gives an overall contribution ${\cal O}(f_{\pi}^{-3} \sqrt{m_\pi /M})$.

One can use similar power-counting arguments to estimate the 
relative contribution of other processes.  In studying other 
processes we have found that they all depend on a higher power of
the pion mass than the $\sqrt{m_\pi}$ dependence of the impulse process.
Thus, power counting verifies the intuition that had we lived in a 
world with sufficiently light quarks the impulse term would have been 
dominant.
Thus, the failure of the impulse term to explain the data by itself is 
interesting in that it highlights the role of chiral-symmetry breaking. 
That is, it quantifies the sense in which the up and down quark 
masses cannot be regarded as sufficiently light.

For example, consider the $\Delta$ terms of Fig.(\ref{f3}).
The power counting here is completely analogous to the case just 
considered;
the only difference is that now the intermediate state energy is
$1/(m_\pi -\delta)$, the rest-mass difference between intermediate 
and initial state being $\delta$, while the typical kinetic energy 
is, as before, of order $m_\pi$. Thus, this process is
${\cal O} (f_{\pi}^{-3} \frac{m_\pi}{(m_\pi -\delta)} \sqrt{m_\pi /M})$ 
or, equivalently, ${\cal O} (\frac{m_\pi}{(m_\pi -\delta)})$
relative to the impulse approximation.
In the limit-world where the quark masses $m_q \rightarrow 0$ implying
$m_\pi \rightarrow 0$, while $m_\Delta - m_N$ remains finite, 
$\Delta$ excitation would be greatly suppressed relative to the 
impulse term. However, in the real (i.e., physical) world, we have
$\frac{m_\pi}{(m_\pi -\delta)} \sim 1$, and this process has to be 
considered as of the same order as the impulse term.

All other processes are suppressed compared to the leading impulse and 
$\Delta$-excitation processes by powers of $\sqrt{m_\pi /M}$.
For example, the recoil corrections to the impulse approximation 
(Fig.(\ref{f4})) and to the static $\Delta$ excitation are 
down by relative order $m_\pi/M$, since they involve two extra factors 
of the ratio between transferred energy ($\sim m_\pi$) and 
transferred momentum ($\sim \sqrt{m_\pi M}$).

The pion-rescattering or seagull process of Fig.(\ref{f5}) 
is also down by relative order $m_\pi/M$ compared to the leading terms.
In order to see this, we first need to understand how the chiral power 
counting goes for the $\pi \pi N^{\dagger} N$ vertex.  
Because of its isospin structure, the Weinberg-Tomozawa term in 
Eq.(\ref{la0}) does not contribute.  
{}From ${\cal L}^{(1)}$ in Eq.(\ref{la1}), we see that 
i) the $c_1$ term and $\delta m_N$, being proportional to the 
$\sigma$-term, are proportional to $m_q$ and hence $m_\pi^2$; 
ii) the $c_2 +c_3$ term yields an interaction  proportional to  
$q_0 E_\pi$, where $q_0$ is the {\it energy} of the exchanged pion, 
so it is also $\propto m_\pi^2$;
iii) the other terms, while potentially big because they are proportional
to the {\it momentum} of the exchanged pion, contribute only to p-waves.
The $\pi \pi NN$ vertex thus contributes a factor $m_\pi^2/(f_\pi^2 M)$.
To complete the power counting for the process, note that
the pion propagator is of order $1/(m_\pi M)$ and 
the coupling of the exchanged pion to the second nucleon is 
$\sqrt{m_\pi M}/f_\pi$.  Combined with the 
seagull vertex, this yields a total amplitude of order 
${\cal O} (f_{\pi}^{-3}(m_\pi/ M)^{3/2})$.

A few comments are in order about the seagull contribution.  The 
first is that although it is formally sub-leading, 
it is by no means obvious that it {\it must} be very small.
Dimensionless spin and isospin factors occur in most amplitudes. They
can significantly enhance some partial waves, and these dimensionless
factors are not ``counted'' in power-counting arguments.
It is also 
clear why this contribution is likely to be much larger 
than estimates based on the on-shell vertex.  The on-shell s-wave 
scattering amplitude is anomalously
small in a power-counting sense.  The basic point is that 
on-shell the $\sigma$ term contribution is nearly equal and 
opposite  to the $c_2$ and $c_3$ contributions, yielding a total 
much smaller than any of the
individual terms.  However, with the kinematics of pion 
production the terms do not cancel so efficiently.
Thus, one expects the on-shell estimates to be low.

Unfortunately, the recoil and the seagull processes are not the 
only ones of this order.  
One-loop diagrams with and without $\Delta$'s 
(e.g., those in Fig.(\ref{f6})) also appear to this order.
Ideally, of course, we should simply calculate these processes
and include them in our analysis.
There are two difficulties with this.
The first is that we are using phenomenological potentials and 
we do not how to implement the traditional  $\chi$PT scheme of 
regulating the loop
and absorbing the regulator dependence into the coefficient of a 
contact term in a way consistent with the potential.
The second is that even if were we able to 
do this in a consistent way, we would not be able to make a 
prediction until we fit the contact term {}from some other process.
Accordingly we have not included these loop diagrams in our analysis.  
We realize, of course, that these graphs could  be quite significant,
especially because there is considerable cancellation among the
other terms.

The most relevant contact terms in question are the ones in Eq.(\ref{la2}), 
and shown in Fig.(\ref{f7}). Their finite parts are expected to be 
${\cal O}(\sqrt{m_\pi M} m_\pi/(f_\pi^3 M m_N))
={\cal O}(f_{\pi}^{-3}(m_\pi/ M)^{3/2})$,
so it is also of the same order as the rescattering mechanism.  
In order to gauge the relevance of this term, we adopt here a
phenomenological strategy: following Ref.\cite{HGM94}, 
we assume it to be saturated by $\sigma$ and $\omega$ exchanges.
If this exchange model is correct, this short-range physics 
is quite significant---the amplitude is of the same magnitude as 
the impulse term.  On the other hand, it is not totally obvious 
that the model is correct.  It depends on using 
large coupling constants and a light $\sigma$.  At least in part, 
this is believed to be a phenomenological parameterization of  
some two-pion-exchange physics in the potential.  
It is not obvious that this phenomenology should be used in a context 
other than the potential.  Accordingly we will first consider
calculations omitting this contribution and subsequently add it 
in assuming the strength given in Ref.\cite{HGM94}. 

We can continue with higher-order examples, but it should be clear now
how this can be generalized. 
(For simplicity we drop here the overall $f_{\pi}^{-3}$ factor.) 
In the Weinberg scheme the amplitude associated with an 
irreducible diagram   is of order $(Q/M)^\nu$ where $\nu$
is given in Eq.(\ref{pc1})
and $Q$ is the typical momentum in the problem or $m_\pi$ or $\delta$,
which are taken to be of the same order.  
With the present kinematics 
the order of an irreducible diagram can be written as
\begin{equation}
A \sim \left (\frac{m_{\pi}}{M} \right )^{\nu_d/2} \left 
(\frac{m_{\pi}}{M} \right )^{\nu_e +\nu_{ex} }
=\left (\frac{m_{\pi}}{M} \right )^{\nu_d/2 + \nu_e + 
\nu_{ex} } \, ,
\end{equation}
where $\nu_{ex}$ counts the order of the explicit chiral symmetry 
breaking arising 
directly {}from the effective Lagrangian, $\nu_{e}$ counts the order 
of the energy transfer in the external legs ({\it i.e.} the total 
number of time derivatives on external legs) and
$\nu_d$ is the dynamical order given by $\nu - \nu_{ex} - \nu_e$. 
 Thus, we can introduce an effective index $\nu_{eff}$ defined by
\begin{equation} 
\nu_{eff} = \nu_d/2 + \nu_e + \nu_{ex}=
(\nu + \nu_e + \nu_{ex})/2 \, ,
\end{equation}
and the irreducible diagram will be of order $(m_{\pi}/M)^{\nu_{eff}}$.

\section{Explicit forms of operators }

We now obtain the explicit forms of the various contributions by
evaluating the most important irreducible diagrams in momentum space.
Our notation is as follows:
$\omega_q^2 = \vec{q}^{\, 2} + m_{\pi}^2$ is the energy of the 
(on-shell)
pion produced with momentum $\vec{q}$ in the center of mass;
$\vec{p}$ ($\vec{p'}$) is the center-of-mass momentum of the incoming 
(outgoing) proton labelled ``1'' (those of proton ``2'' are opposite); 
$\vec{k}= \vec{p} -\vec{p'}$ ($k^0= (\vec{p}^{\, 2} -\vec{p'}^2)/2m_{N}$)
is the momentum (energy) transferred;
$\omega_k^2 = \vec{k}^2 + m_{\pi}^2$;
$\vec{P}= \vec{p} +\vec{p'}$;  
$\vec{\sigma}^{(i)}$ is the spin of proton $i$;
$\vec{\Sigma}= \vec{\sigma}^{(1)} - \vec{\sigma}^{(2)}$; 
and $T(\vec{k})\equiv \vec{\sigma}^{(1)}\cdot\vec{k} \,
\vec{\sigma}^{(2)}\cdot\vec{k}$.  We define the T-matrix in terms of the 
S-matrix via S=1+$i$T. 

According to the previous discussion, we expect the leading contributions
to arise {}from the diagrams in Figs.(\ref{f2}) and (\ref{f3}) where the
virtual pion is exchanged between static baryons. 
In the case of pion exchange with a nucleon in the intermediate state 
(Fig.({\ref{f2}a,b)), we get
\begin{equation}
T^{IA,p}=\frac{ig_A^3}{8 m_{N} f_{\pi}^3} \frac{1}{\omega_k^2}
       \left[\vec{\Sigma}\cdot\vec{p'} T(\vec{k})- 
        T(\vec{k}) \vec{\Sigma}\cdot\vec{p}\right]\, ,
\label{first}
\end{equation}
\noindent which is listed for comparative purposes only. We will actually
calculate the impulse approximation directly from Eq.(6).
Recoil corrections, as in Fig.(\ref{f4}),  are expected to be smaller
by a factor of $m_{\pi}/M$, and are also included:
\begin{equation}
T^{recoil}=\frac{ig_A^3}{8 m_{N} f_{\pi}^3} \frac{(k^0)^2}{\omega_k^2}
       \frac{1}{\omega_k^2 -(k^0)^2}
       \left[\vec{\Sigma}\cdot\vec{p'} T(\vec{k})- 
        T(\vec{k}) \vec{\Sigma}\cdot\vec{p}\right].
\label{second}
\end{equation}
\noindent 
In the case of pion exchange with a delta intermediate state 
(Fig.(\ref{f3}a,b)), 
\begin{eqnarray}
T^{\Delta} & = & 
       \frac{-ig_A h_A^2}{18 m_{N} f_{\pi}^3} \frac{1}{\omega_k^2-(k^0)^2}
       \frac{\omega_q}{\delta^2-\omega_q^2}   \nonumber \\
  & &   \left[(\vec{k}^2 \omega_q - \vec{k}\cdot\vec{P}\delta)   
        \vec{\Sigma}\cdot \vec{k}
       +\frac{i}{2} \omega_q (\vec{\sigma}^{(1)}\cdot\vec{k} 
        \vec{\sigma}^{(2)}\cdot\vec{P}\times\vec{k}           
        -\vec{\sigma}^{(1)}\cdot\vec{P}\times\vec{k}         
        \vec{\sigma}^{(2)}\cdot\vec{k})\right].
\label{third}
\end{eqnarray}
\noindent
Results similar to Eqs.(\ref{first}),(\ref{third}) follow for the 
shorter-range terms (Fig.(\ref{f2}c,d), Fig.(\ref{f3}c,d)), 
and we do not write them explicitly here. 
The diagrams in Fig.(\ref{f2}) will all be included in Fig.(\ref{f1}) 
as far as the explicit calculation goes.  
Diagrams in Fig.(\ref{f3}c,d) have to be accounted
for explicitly as well as diagrams in Fig.(\ref{f3}a,b). 
In any reasonable model, however,
they turn out to be smaller than those in diagrams of Fig.(\ref{f3}a,b). 
For example,
they could arise {}from $a_1$ exchange, but then the relatively high
$a_1$ mass suppresses this contribution. 
For the purpose of estimating the effect of the $\Delta$, we use 
Eq.(\ref{third}).

There are two other corrections of order $m_{\pi}/M$ compared to the
leading diagrams of Figs.(\ref{f2}) and (\ref{f3}).
Fig.(\ref{f5}) represents the s-wave rescattering:
\begin{equation}
T^{ST}=\frac{ig_A}{f_{\pi}^3} \frac{1}{\omega_k^2 -(k^0)^2}
       \left[(c_2 +c_3- \frac{g_A^2}{8 m_N}) k^0 \omega_q
    -2c_1 m_{\pi}^2 -\frac{\delta m_N}{4}\right] \vec{\Sigma}\cdot\vec{k}.
\end{equation} 
\noindent
Fig.(\ref{f7}) is a short-range mechanism provided by Eq.(\ref{la2}):
\begin{equation}
T^{sr}= \frac{i}{2 f_{\pi} m_N} \omega_q 
        \left[(d_1^{\prime}+2e_1) \vec{\Sigma}\cdot\vec{P} 
     + 2 i e_2 \vec{\sigma}^{(1)}\times\vec{\sigma}^{(2)}\cdot\vec{k}\right] 
       + \ldots.
\label{dande}
\end{equation}
\noindent
Chiral symmetry tell us nothing about the strength of these terms.
We can use data to determine the coefficients $d_1^{\prime}$, $e_1$, and $e_2$. 
Alternatively,
we can use a model to determine these coefficients and then try to explain
the experimental results. Here we use the mechanism first proposed
by Lee and Riska \cite{LR92} and by Horowitz {\it et al.} \cite{HGM94},
where the short-range interaction is supposed to originate {}from
z-graphs with $\sigma$ and $\omega$ exchanges. In this case,
\begin{eqnarray}
T^{\sigma, \omega} & = & - \frac{i\, g_A}{4 f_{\pi} m_N^2} \omega_q 
      [\left(\frac{g_\sigma^2}{\vec{k}^2 + m_{\sigma}^2}+ 
                   \frac{g_\omega^2}{\vec{k}^2 + m_{\omega}^2}\right)
              \vec{\Sigma}\cdot\vec{P} \nonumber\\ 
  &   & -2 i\frac{g_\omega^2 (1+C_\omega)}{\vec{k}^2 + m_{\omega}^2} 
         \vec{\sigma}^{(1)}\times\vec{\sigma}^{(2)}\cdot\vec{k}]
    \, + \cdots ,
\label{last}
\end{eqnarray}
where $m_\sigma$ ($m_\omega$) and $g_\sigma$ ($g_\omega$) are 
the mass and the (vector) coupling to nucleons of the 
$\sigma$ ($\omega$) meson, 
and $C_\omega$ denotes the ratio of tensor to vector coupling for the 
$\omega$ meson. (Note that there is a misprint in the sign of the second term
 of Eq. (7b) of Ref.\cite{LR92}; the $i$ should be replaced by $-i$.)
  Comparison between Eqs.(\ref{dande}) and (\ref{last}) 
yields the  
{\it model-dependent} estimates: 
$d_1^{\prime} + e_1 = -g_A g_\sigma^2 /2 m_{\sigma}^2 m_N \approx
-1.5 (1/f_{\pi}^2 M)$,  
$e_1 = -g_A g_\omega^2 /2 m_{\omega}^2 m_N \approx -2 (1/f_{\pi}^2 M)$, and
$e_2 = g_A g_\omega^2 
(1+C_\omega)/2 m_{\omega}^2 m_N \approx 2 (1/f_{\pi}^2 M)$. 
The numerical factors are obtained from the Bonn OBEP A potential, Table A.2
of reference \cite{machleidt}.

We are concerned with evaluating the matrix elements of the above 
operators between the initial $^3P_0$ and final $^1S_0$ 
$pp$ wave functions.
It is convenient to use the Reid\cite{Reid}, Reid93 \cite{Reid93}
and Argonne V18 \cite{v18} potentials which, for a given $pp$ channel, 
are local potentials. Thus we evaluate the operators between coordinate 
space initial ($i$) and final ($f$) 
wave functions expressed by
\begin{equation}
\langle \vec{r}|i \rangle ={\sqrt{2}\over pr} i \, u_{1,0}(r)e^{i\delta_{1,0}} 
\sqrt{4\pi}\, |^3P_0\rangle,
\end{equation}
and
\begin{equation}
\langle \vec{r}|f \rangle ={1\over p'r} u_{0}(r)e^{-i\delta_{0}} 
\sqrt{4\pi}\, |^1S_0\rangle\, .
\end{equation}
We convert the operators of Eqs.(12)-(\ref{last}) to
configuration space by inverting the Fourier transforms.
The resulting operators can then be used in configuration-space matrix
elements. Hermiticity and $[\vec\sigma^{(1)}\cdot \vec\sigma^{(2)},
\vec\Sigma\cdot \hat r] = 4 i \vec\sigma^{(1)}\times \vec\sigma^{(2)} 
\cdot \hat r$ lead to
\begin{eqnarray}
\langle ^1S_0|i \vec\sigma^{(1)}\times \vec\sigma^{(2)} \cdot 
\hat r |^3P_0\rangle = - \langle ^1S_0|\vec\Sigma\cdot \hat r |^3P_0\rangle 
\, ,\nonumber\\
\langle ^1S_0|\vec\Sigma\cdot \vec p|^3P_0\rangle = 
-i \langle ^1S_0|\vec\Sigma\cdot \hat r |^3P_0\rangle
\left({\partial\over\partial r}+{2\over r}\right)\, ,
\end{eqnarray}
while direct evaluation leads to
\begin{equation}
\langle ^1S_0|\vec\Sigma\cdot \hat r |^3P_0\rangle = -2\, .
\end{equation}

We may then begin to tabulate the results. We define the matrix 
elements of the operators of Eqs.(12)-(\ref{last}) as  
\begin{eqnarray}
{\cal M}^{X} & = & \langle f|T^{X}|i\rangle  \nonumber \\
      & = & -{4\pi\sqrt{2}\;i\over p p'}
            e^{i (\delta_0 + \delta_{1,0})} 
            \int_0^\infty dr\;u_0(r)\,H^{X}(r)\,u_{1,0}(r),
\label{Mdef}
\end{eqnarray}
\noindent
where $X$ represents $IA$, $recoil$, etc, and $H^{X}(r)$ is the
corresponding operator, to be given shortly. We have extracted all of the 
constant factors from Eqns.(17) and (18) together with a sign from 
Eqns.(19) and (20) and placed them in front of the integral.
The integral can be broken up into a part {}from zero to $r_0$ where
$r_0$ is any distance greater than the range of the nucleon-nucleon 
force. For $r>r_0$, the wave functions $u_0, u_{1,0}$ are linear 
combinations of regular and irregular Coulomb wave functions, 
so that the integral {}from $r_0$ to $\infty$ can be done analytically.
Note that we can also define $J$-matrix elements $J^X$, 
so that $J^{IA}=J_1$  of Ref.~\cite{KR66}. The  relationship is given by
\begin{equation}
{\cal M}^{X} \equiv 
i\sqrt{2}{4\pi\over pp^\prime}{g_A\over f_\pi}{1\over m_Nm_\pi}J^{X}.
\label{tdef}
\end{equation}
The impulse approximation is given by
\begin{equation}
H^{IA}(r)=-\frac{g_A}{f_\pi} {m_\pi \over m_N}
           \left({\partial\over\partial r}+{1\over r}\right).
\label{ia}
\end{equation}
The recoil term is evaluated {}from
\begin{equation}
H^{recoil}(r)=\frac{g_A}{f_\pi} {1\over m_N}\left[4{\partial\over \partial\;r}
\left(V_T(\tilde{m}_\pi,r)-V_T(m_\pi,r)\right)-
{\partial\over \partial\;r}
\left(V_S(\tilde{m}_\pi,r)-V_S(m_\pi,r)\right)\right],
\label{recoil}
\end{equation}
with
\begin{equation}
\tilde{m}_\pi \equiv\sqrt{\frac{3}{4}} m_\pi,
\end{equation}
and
\begin{eqnarray}
V_S(\mu,r)&=&({g_A\over 2f_\pi})^2 {\mu^3\over 4\pi}{e^{-\mu\;r}\over
\mu\;r}\, , \\
V_T(\mu,r)&=&({g_A\over 2f_\pi})^2 {\mu^3\over 4\pi}{e^{-\mu\;r}\over
\mu\;r}\left({1\over \mu^2\;r^2}+{1\over \mu\;r}+{1\over 3}\right)\, .
\end{eqnarray}
The dependence on $\tilde{m}_\pi$ arises {}from the restriction occurring 
at threshold that the total momentum of the two final nucleons be zero.
In that case, we have $k^0\to m_\pi/2$.

The seagull term ($ST$) is evaluated in a similar manner {}from
\begin{equation}
H^{ST}(r)=-{g_A m_{\pi}^2\over f^3_\pi}\left[4c_1+{\delta m_N\over 2 m_\pi^2}
-\left(c_2+c_3-{g_A^2\over8m_N}\right)\right]{(1+\tilde{m}_\pi r)\over
4\pi\;r^2} e^{-\tilde{m}_\pi r}.
\label{stoff}
\end{equation}
\noindent
One may also evaluate the  graph of Fig.(\ref{f5}) by treating the 
pion-nucleon vertex in an on-shell approximation. In this case the 
term $k^0$ is replaced by $m_\pi$. The result is that one obtains 
an on-shell form of the seagull term ($STon$) with
\begin{equation}
H^{STon}(r)=-{g_A m_{\pi}^2 \over f^3_\pi}\left[4c_1
+{\delta m_N\over 2 m_\pi^2}
-2\left(c_2+c_3-{g_A^2\over8m_N}\right)\right]{(1+\tilde{m}_\pi r)\over
4\pi\;r^2} e^{-\tilde{m}_\pi r}.
\label{ston}
\end{equation}

The term involving the intermediate $\Delta$ is found to take the form
\begin{equation}
H^{\Delta}(r)={g_A^3\over 18m_N\,f_\pi^3}
            \left(h_A\over g_A\right)^2 {m_\pi^2
\tilde{m}_\pi^2\over\;m_\pi^2-\delta^2}
[-A_\Delta(r) +(1-{2\delta \over m_\pi}) B_\Delta(r)
      +C_\Delta(r) ],
\label{delta}
\end{equation}
where 
\begin{eqnarray}
A_\Delta(r)&=&{3(1+\tilde{m}_\pi\;r)\over 4\pi\;r^2}e^{-\tilde{m}_\pi\;r},
\nonumber\\
B_\Delta(r)&=&{e^{-\tilde{m}_\pi\;r}\over4\pi\;r}[
  {(1+\tilde{m}_\pi\;r)\over \;r}
+{2\over \;r}\left(1+{4\over \tilde{m}_\pi\;r}
+{4\over \tilde{m}_\pi^2\;r^2}\right)
-2\left(1+{2\over \tilde{m}_\pi\;r}
+{2\over \tilde{m}_\pi^2\;r^2}\right){\partial\over \partial r}], \\
C_\Delta(r)&=&{2\over 4\pi}{e^{-\tilde{m}_\pi\;r}\over r}
[{\partial\over \partial r}
+{1\over r}]. \nonumber
\end{eqnarray}

We also include the effects of the sigma and omega exchange terms 
($\sigma, \omega$). 
The result is 
\begin{equation}
H^{\sigma, \omega}(r)=-\frac{g_A}{f_\pi} {m_\pi \over m_N^2}
              \left[ (f_\sigma(r)+ f_\omega(r))
              \left({\partial\over\partial r}+{1\over r}\right)
              + \frac{1}{2}\frac{\partial f_\sigma}{\partial r}
              - \frac{1+2 C_\omega}{2}\frac{\partial f_\omega}{\partial r}
\right]\, ,
\label{sigma}
\end{equation}
\noindent
where the function $f_h(r)$ accounts for exchange of the meson $h$ between 
nucleons,
\begin{equation}
f_h(r)={g_h^2\over 4\pi}{e^{-m_h r}\over r}.
\end {equation}
We follow Ref.\cite{HGM94} by including the effects of form factors as 
defined in the Bonn potential \cite{machleidt}. 
For completeness, we repeat that monopole form factors are used 
at each meson vertex according to the replacement
\begin{equation}
g_h\to g_h{\Lambda_h^2-m_h^2\over\Lambda_h^2-k_\mu k^\mu},
\label{formf}
\end{equation}
where $k_\mu$ is the transferred momentum and $\Lambda_h$ is the 
cutoff mass.

These expressions for $H^{\sigma,\omega}$ differ from the 
corresponding ones of Ref.\cite{HGM94} because we use $\vec{p}^{\, \prime}
=\vec p -\vec k$ instead of approximating $\vec{p}^{\, \prime}$ by $\vec p$ 
as in the earlier work. 
The failure of this approximation
has very recently been pointed out by Niskanen\cite{JN96}.
Keeping the 
term proportional to $\vec k$ yields terms proportional to 
${df_{\sigma,\omega}\over dr}$ that are not included in 
Ref.\cite{HGM94}. However, this change does not effect the conclusions of 
Ref.\cite{HGM94}, because a decrease in the contribution of the sigma exchange
 is
compensated by an increase of that of the omega exchange.

The final steps consist of computing the total matrix element ${\cal M}$:
\begin{equation}
{\cal M}={\cal M}^{IA}+{\cal M}^{ST}+{\cal M}^{recoil}+{\cal M}^\Delta
+{\cal M}^{\sigma, \omega},
\label{calm}
\end{equation}
squaring that sum, and integrating over the available phase space. 
We find
\begin{equation}
\sigma={1\over v}\int_0^{p^\prime_{max}} {dp'p'^2\;q'\over (2\pi)^3}|{\cal M}|^2
{m_N\over 2m_N+\omega(q')},
\label{cross}
\end{equation}
where $v$ is the laboratory velocity of the incident proton, 
$q'$ is the pion momentum, $\omega(q')=\sqrt{q'^2+m_\pi^2}$ and
$p^\prime_{max}=\sqrt{p^2-m_Nm_\pi}$.

\section{Input Parameters and Results}

The various amplitudes considered in the last section depend on
several parameters that we can determine {}from other processes.
The impulse-approximation operator of Eq.(\ref{ia}) and the recoil
operator of Eq.(\ref{recoil}) depend on 
the pion mass, $m_\pi = 135$ MeV \cite{pdb}, and on   
\begin{equation}
\frac{g_A}{f_{\pi}}=\frac{g_{\pi NN}}{m_N};
\end{equation}
\noindent
we use the value of $g_{\pi NN}$ appropriate for each potential. 
The $\Delta$ operator of Eq.(\ref{delta}) further depends 
on the $\Delta-N$ mass splitting $\delta = 294$ MeV \cite{pdb} and  
on the $\pi N \Delta$
coupling constant, $h_A$. This has been fixed {}from p-wave
$\pi N$ scattering (see, e.g. Ref.\cite{weise}), 
\begin{equation}                             
\frac{h_A}{g_A} \simeq 2.1.
\end{equation}
\noindent
The seagull operator of Eq.(\ref{stoff}) depends on four parameters
$c_{1,2,3}$ and $\delta m_{N}$. 
The $c_{i}$'s can be obtained by fitting s-wave $\pi N$ scattering.
In Ref.\cite{BKM95} they were found to be
\begin{eqnarray}
c_1 &=& -1.63/2m_N\nonumber\\
c_2 &=& 6.20/2m_N\label{ceq}\\
c_3 &=& -9.86/2m_N\nonumber \, ,
\end{eqnarray}
{}from the $\sigma$-term, the isospin-even scattering length, 
and the axial polarizability.
Note that the analysis of Ref.\cite{BKM95} does not include 
the isobar explicitly. 
Since the inclusion of the $\pi N \Delta$ interaction  
only affects s-waves at one order higher than the $c_{i}$'s,
the above values can still be used to estimate the effect of
s-wave rescattering.  
The remaining parameter, $\delta m_{N}$, can in principle also
be determined {}from s-wave $\pi N$ scattering, but would require
a careful analysis of other isospin-violating effects. Chiral
symmetry relates it to the
strong interaction contribution to the nucleon mass splitting, which
is also difficult to determine directly. 
Estimates of the electromagnetic contribution $\bar{\delta} m_{N}$
are more reliable, $\bar{\delta} m_{N}\sim -1.5$ MeV \cite{jerry},
and give $\delta m_{N} \sim 3$ MeV. To be definite, we use
\begin{equation}
\delta m_{N}=3 \, \mbox{MeV}.
\label{deltmn}
\end{equation}
Finally, the $\sigma, \omega$ operator of Eq.(\ref{sigma}) involves 
$g_{h}$, $\Lambda_{h}$, $m_{h}$, and $C_\omega$, parameters listed in
Table A.3 of Ref.\cite{machleidt}.

We shall consider results of individual terms before presenting 
complete calculations using $\cal M$ of Eq.(\ref{calm}). 
We shall often compare our results with the IUCF \cite{M90}
and Uppsala \cite{uppsala} data.
This allows us to understand whether or not the size of a particular 
term is relevant. The values of  the cross sections divided by 
$\eta^2$ will be displayed, where $\eta$ is the maximum value 
of the pion momentum divided by the mass of the $\pi^0$. 
The three points at the lowest values of $\eta$ are
{}from Ref.~\cite{uppsala}, the remainder  are {}from Ref.~\cite{M90}.
We restrict our calculations to values of $\eta$ such that $\eta 
\le 0.4$, because the $\vec\sigma\cdot \vec{\nabla} \boldpi$ term 
of the Lagrangian (which we ignore) is important for higher 
values \cite{julich}.

We start by considering the effects of the impulse-approximation term
of Eq.(\ref{ia}). We compute the cross section using 
${\cal M}={\cal M}^{IA}$.
The results are shown in Fig.(\ref{ianoc}), where the Coulomb 
distortion of the proton-proton initial and final wave functions 
is neglected, and in Fig.(\ref{iawic}), where the Coulomb effects 
are included. We observe that the impulse approximation provides  
cross sections much smaller than the measured ones, and that including 
Coulomb effects is important in describing the $\eta$ dependence of 
the new accurate cross section data. Henceforth, the only results we 
shall present include the Coulomb effects.

The observations about the small size of the impulse term and the 
importance of the Coulomb distortion 
have been made before\cite{MS91,M90,HGM94}. 
Furthermore, the small values of the cross sections can be traced to 
cancellations in the integrands. This is shown in 
Fig.(\ref{iacoord}). The oscillations arise {}from the high momentum 
($p \sim \sqrt{m_N\;m_\pi}$)
of the initial state.

Thus terms other than the impulse approximation must be included.
This is also the case in the $pp\to d\pi^+$ reaction and in pion 
production on nuclear targets (e.g., see the reviews \cite{M79,F80}). 
The computation of the terms of Eqs.(\ref{Mdef}) and 
(\ref{recoil})-(\ref{formf}) is straightforward. These matrix 
elements are evaluated as a function of $p^\prime$  
and for $\eta=0.3$ using the Reid potential (Fig.(\ref{mat3})), 
the Reid93 potential (Fig.(\ref{mat3r93})) and the V18 potential
(Fig.(\ref{mat3v18})). The $pp\to d\pi^+$
cross sections can be understood in terms of rescattering mechanisms, 
so it is natural to include the effects of the rescattering via the 
seagull term. We see that including this term in a manner 
dictated by the chiral Lagrangian (Eq.(\ref{stoff})) leads to a 
matrix element with a sign opposite to that of the impulse-approximation 
term. These two terms cancel to a large extent, so that one is forced to 
examine other terms. If one uses on-shell kinematics to evaluate the 
influence of the  seagull term (Eq.(\ref{ston})), 
one finds a contribution of the same sign as the impulse term.  

The reason the seagull amplitude changes its sign
when going {}from on-shell to half-off-shell kinematics can be seen 
immediately by comparing the square brackets of Eqs.(\ref{stoff}) 
and (\ref{ston}) using the values of $c_i$ of Eq.(\ref{ceq}) and 
$\delta m_N$ Eq.(\ref{deltmn}).
One finds for the quantity governing off-shell rescattering 
\begin{equation}
\left[4c_1+{\delta m_N\over 2 m_\pi^2}
-\left(c_2+c_3-{g_A^2\over8m_N}\right)\right]
= -{2.31\over 2\;m_N}
\end{equation}
and for the quantity governing the on-shell rescattering
\begin{equation}
\left[4c_1+{\delta m_N\over 2 m_\pi^2}
-2\left(c_2+c_3-{g_A^2\over8m_N}\right)\right]= +{1.75\over2\;m_N}.
\end{equation}
\noindent            
Note that the contribution {}from isospin violation 
($-\delta m_N /2 m_\pi^2$) is about 10\% of the total on-shell 
contribution, which is relatively large but insufficient to play 
a significant role in the current stage
of understanding of the $pp \rightarrow pp \pi^0$ reaction.

We also discuss the other terms.
The recoil term should be of order ${m_\pi /M}$ relative to the 
impulse term, and it is indeed quite small, as indicated in the 
figures. The term involving the intermediate  $\Delta$ is also 
smaller than impulse, but depending on the potential is about 
2--4 times bigger than the recoil term.
The smaller nature of the term results {}from cancellations in 
the integrand, as shown in Fig.(\ref{xxdelta.top}). 
The sensitivity of the value of the matrix element 
to the choice of potential is also due to this cancellation. 
Qualitatively the integrands shown in the figure are very 
similar, but the net values shown in Figs.(\ref{mat3}) and 
(\ref{mat3v18}) are very different.

The importance of the interference between the impulse and 
seagull terms is underscored in Fig.(\ref{tot}), 
which shows the result of the sum of 
the impulse, recoil, seagull and $\Delta$ terms for the
three different potentials that we use. 
The correct evaluation of our Lagrangian leads to cross sections 
that are very nearly zero! Only the lowest-energy data point can be 
shown on  this figure with a linear scale. 
Including the effects of the $\Delta$ in the intermediate state is 
not sufficient to provide a description of the data.
None of the calculated cross 
sections is near the data, except at the lowest energy.

This failure to account for the measured cross sections causes us to
examine the effects of also including the effects of $\sigma$ and
$\omega$
exchange. 
As shown in Fig.(\ref{sigmat}) we can not account for the 
cross sections with the heavy meson exchange
effects. This is because the $\Delta$ and pion seagull terms interfere
 destructively with
the other terms.

\section{Discussion}  

The cross sections we compute are far smaller than the measured 
ones, so we cannot account for this process in terms of the physics 
we have included.  However, we have learned something significant 
about the reaction.  We have found processes that contribute amplitudes  
of a similar size to the impulse process and which should be included 
in any reasonable description: the $\Delta$ process, and the seagull 
or pion-rescattering process.  
The $\Delta$ process has not been considered previously for the 
$pp \to pp \pi^0$ reaction at threshold, but it is important when 
the newest $pp$ potentials are used to describe the scattering wave 
functions.  We have also seen that the seagull process is 
highly dependent on its off-shell behavior.

The validity of the previous conclusions is independent of chiral 
perturbation theory itself (i.e., of the convergence of the expansion 
in momenta).
The relevance of the $\Delta$ isobar depends mostly on the observations
i) that it is a relatively light degree of freedom and
ii) that its coupling to a nucleon and an s-wave pion is essentially
fixed by Galilean invariance in terms of its (large) coupling to a p-wave 
pion.
The importance of the off-shell behavior of the s-wave $\pi^0 N$ 
(re)scattering results {}from the existence of momentum-dependent
$\pi \pi NN$ interactions of a size similar to the momentum-independent
one (which is proportional to the quark masses). This in turn 
is a consequence of chiral symmetry.
  
We have also seen that these two processes interfere destructively
with the impulse term. 
This is in contrast with the recent calculation of Ref.\cite{HO95},
where constructive interference was found between impulse and seagull 
terms.  
It is useful to understand the origin of this discrepancy.  

The key point is that the magnitude and sign of the seagull
process do depend on our power-counting arguments.
So far as the  power counting is concerned, we have identified an 
interesting problem: how one proceeds when the typical momenta are 
$\sim (m_\pi m_N)^{1/2}$.
This is relatively straightforward with the choices of pion and 
nucleon fields in which the pseudo-Goldstone-boson nature of the 
pion is manifest;
in any such basis, as used above, these fields provide non-linear 
realizations of chiral symmetry.
As is well--known, it is possible to redefine these fields
without changing the physics, provided we use our field definition 
consistently throughout the calculation \cite{CWZ}. 
Different choices of interpolating fields amount, after all, 
to no more than different bookkeeping of where 
various bits of the physics reside. 
For example \cite{jim}, one can transform to
a linear basis at the expense of producing extra seagull vertices
that are momentum independent but {\it not} proportional to the quark
masses. The sole purpose of such large terms is to ensure a cancellation
of large pole terms and to produce a small
$\pi N$ scattering amplitude. Such a cancellation would seem magical 
were it not understood as a consequence of a concealed chiral symmetry; 
power counting in this basis becomes highly non-trivial. 
Moreover, failure to include such seagull vertices in a calculation
would spoil chiral symmetry.

Now, the pion field in Ref.\cite{HO95} and in $\chi$PT are not equivalent.  
The pion interpolating field in Ref.\cite{HO95} is based on models 
exploiting PCAC whereas the pion field in the chiral Lagrangian is not.  
Thus, {\it a priori} there is nothing wrong with getting a different 
result for the seagull term contribution in the two calculations.  
If both calculations consistently included all other processes,
then the differences between the present calculation and 
Ref. \cite{HO95} in the off-shell seagull contributions would 
be compensated for by differences in the contributions of other 
mechanisms.
Indeed, starting {}from our general chiral Lagrangian and performing 
a simple pion-field redefinition, we can reproduce the off-shell 
behavior of the PCAC $\pi N$ amplitude of Ref.\cite{HO95}; 
however, in doing so an extra, anomalously large contact term is 
generated, whose sole function is to cancel the effect of 
the change in off-shell behavior of the $\pi N$ amplitude
brought up by the field redefinition. This is in complete analogy 
to the change to a linear basis mentioned above.                             
Since the calculation of Ref.\cite{HO95} does not include other 
mechanisms besides the impulse and seagull terms,
it is rather difficult to interpret its results.

We are not able to reproduce the data with the processes we have 
included, even if
the short-ranged process is modeled by $\sigma$ and $\omega$
exchange using the large coupling constants of one boson exchange
potentials. 
Perhaps, in face of the preceding remarks, this should come
as no surprise. To the extent that pieces in the potential
other than static OPE are important, we might expect similarly
important contributions {}from one loop diagrams 
(such as the ones in Fig.(\ref{f6})) in the irreducible
sub-diagram, again to be evaluated consistently
with the potential. 
It might be no accident, then, that without such contributions 
we can not explain the data.

The threshold behavior of $p p \rightarrow p p \pi^0$ in principle 
provides us with considerable information about the nature of 
low-energy strong-interaction  physics and in particular about the 
dynamics associated with approximate chiral symmetry and its 
spontaneous breaking.
The amplitude for this process would be zero in the chiral limit 
of $m_q =0$.  Moreover,  as demonstrated by the power-counting 
analysis in this paper, we see that for sufficiently small $m_q$ the 
amplitude must be dominated by the impulse term. However, we have also
shown that the chiral expansion for a process with these kinematics is
essentially an expansion in $(m_\pi/M)^{1/2}$.  In the real world, 
this parameter is rather large, $\sim .4$, and it is by no means 
surprising that the expansion either converges 
slowly or fails as an asymptotic series after some low order.
Given this large effective expansion parameter, it is not surprising
that many processes contribute non-negligible amounts to the final
amplitude.  Moreover, there is significant cancellation between 
amplitudes {}from the various mechanisms which makes  reliable
calculations very difficult. However, even given the large 
uncertainties present in our calculations due to these cancellations, 
we cannot explain the data given the processes we have included.  
It is reasonably plausible that the terms we have excluded could
provide a large enough contribution to bring our calculations into 
line with the observations.  
Clearly, however, developing a reliable systematic description of 
this processes which explains the data represents a large challenge.

\vspace{2cm}
{\large \bf Acknowledgements} 

We thank M. Savage and  D. Kaplan for useful discussions.  
We also thank C.J. Horowitz, T.-S.H. Lee, J.A. Niskanen, and D.O. Riska
for discussions about their heavy meson theory.   TDC 
and JLF thank the national INT for its gracious hospitality.
This research was supported in part by the Department of Energy
grant DE-FG06-88ER40427.  TDC also acknowledges the support of the U.S. 
Department of Energy under grant  DE-FG06-93ER-40762 and the National
Science Foundation  under grant PHY-9058487. The work of JLF was performed 
under the auspices of the Department of Energy.

\newpage

\begin{figure}
\caption{Various contributions to the $p p \rightarrow p p \pi^0$ reaction.
In this and the following figures a  single (double) solid line stands
for a nucleon (Delta) and a single (double) dashed line represents a
pion (sigma, omega); $\Psi_i$ ($\Psi_f$) is the wave function for the 
initial (final) state.}
\label{f1}
\end{figure}

\begin{figure}
\caption{Diagrams contributing to the impulse term that are
irreducible in the context of chiral power counting for the
$p p \rightarrow p p \pi^0$ reaction.}
\label{f2}
\end{figure}

\begin{figure}
\caption{Irreducible diagrams contributing to the leading 
$\Delta$-excitation mechanism.}
\label{f3}
\end{figure}

\begin{figure}
\caption{Irreducible diagrams contributing to the recoil correction 
of the impulse term.}
\label{f4}
\end{figure}

\begin{figure}
\caption{The seagull or pion rescattering irreducible diagram.}
\label{f5}
\end{figure}

\begin{figure}
\caption{Some one loop irreducible diagrams.}
\label{f6}
\end{figure}

\begin{figure}
\caption{Short-ranged irreducible diagram.}
\label{f7}
\end{figure}

\begin{figure}
\caption {Impulse approximation. Three different potentials are used.
The Coulomb distortion is neglected.
The solid curve is obtained using the 
Reid potential, the dashed curve using  the Reid93 potential 
and the dot-dashed curve using
the V18 potential.}
\label{ianoc}
\end{figure}

\begin{figure}
\caption {Impulse approximation. Three different potentials are used.
The Coulomb distortion is included. The solid curve is obtained using the 
Reid potential, the dashed curve using  the Reid93 potential 
and the dot-dashed curve using
the V18 potential.}
\label{iawic}
\end{figure}

\begin{figure}
\caption {Typical integrand for the impulse term.
The dashed curve is obtained using the Reid potential and the solid with the 
V18 potential.}
\label{iacoord}
\end{figure}

\begin{figure}
\caption {Matrix elements as a function of $p^\prime$ for $\eta=0.3$. 
See Eq.(22). 
The Reid potential is used.} 
\label{mat3}
\end{figure}

\begin{figure}
\caption {Matrix elements as a function of $p^\prime$ for $\eta=0.3$. 
See Eq.(22). 
The Reid93 potential is used.}
\label{mat3r93}
\end{figure}

\begin{figure}
\caption {Matrix elements as a function of $p^\prime$ for $\eta=0.3$. 
See Eq.(22). 
The V18 potential is used.}
\label{mat3v18}
\end{figure}
                                               
\begin{figure}
\caption {Integrands for the $\Delta$ contribution.
The solid curve is obtained with the Reid
and the dashed with the V18 potential.}                            
\label{xxdelta.top}
\end{figure}

\begin{figure}
\caption {${\cal M}\approx{\cal M}^{IA}+{\cal M}^{recoil}
+{\cal M}^{ST}+{\cal M}^\Delta$.}
 \label{tot}
\end{figure}

\begin{figure}
\caption {Calculations with all of our mechanisms for 
the Reid, Reid93, and V18 potentials, using eq. (32) for the 
heavy meson exchange effects.}
\label{sigmat}
\end{figure}

\newpage

\begin{figure}
\centerline{\epsffile{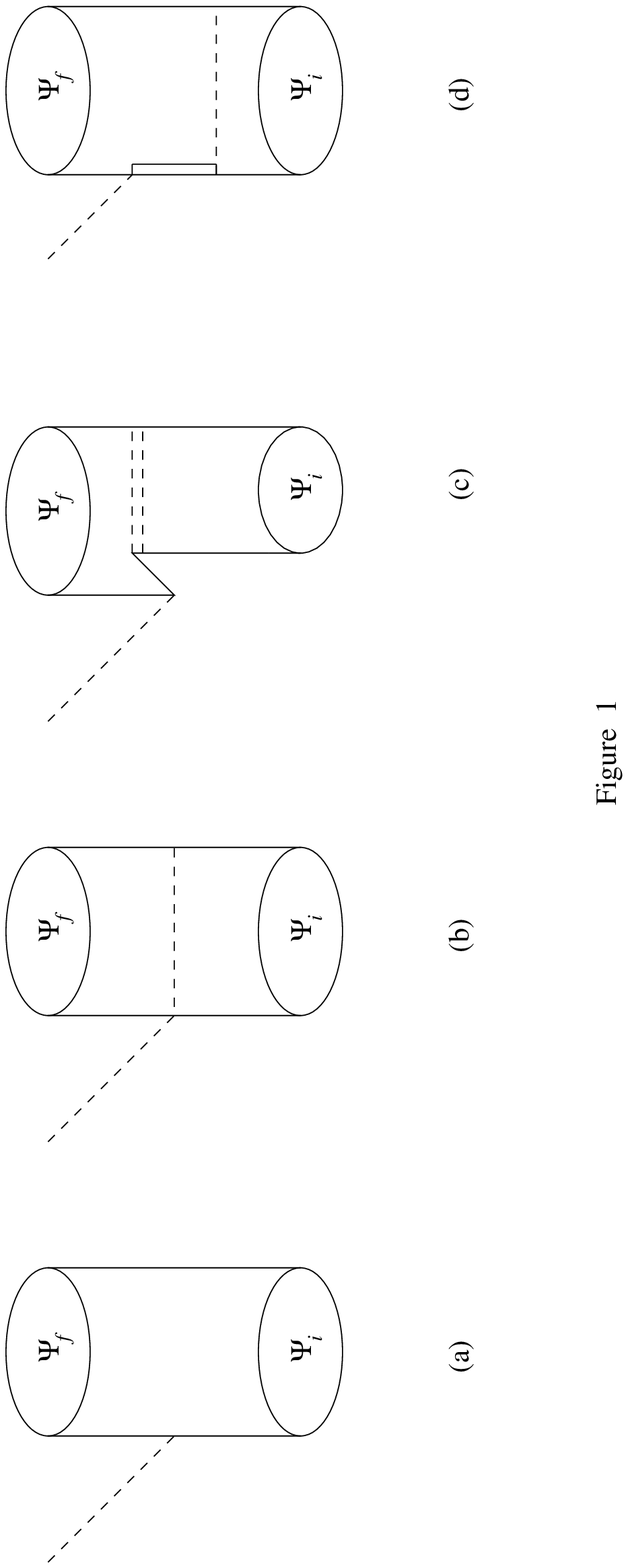}}
\end{figure}

\newpage

\begin{figure}
\centerline{\epsffile{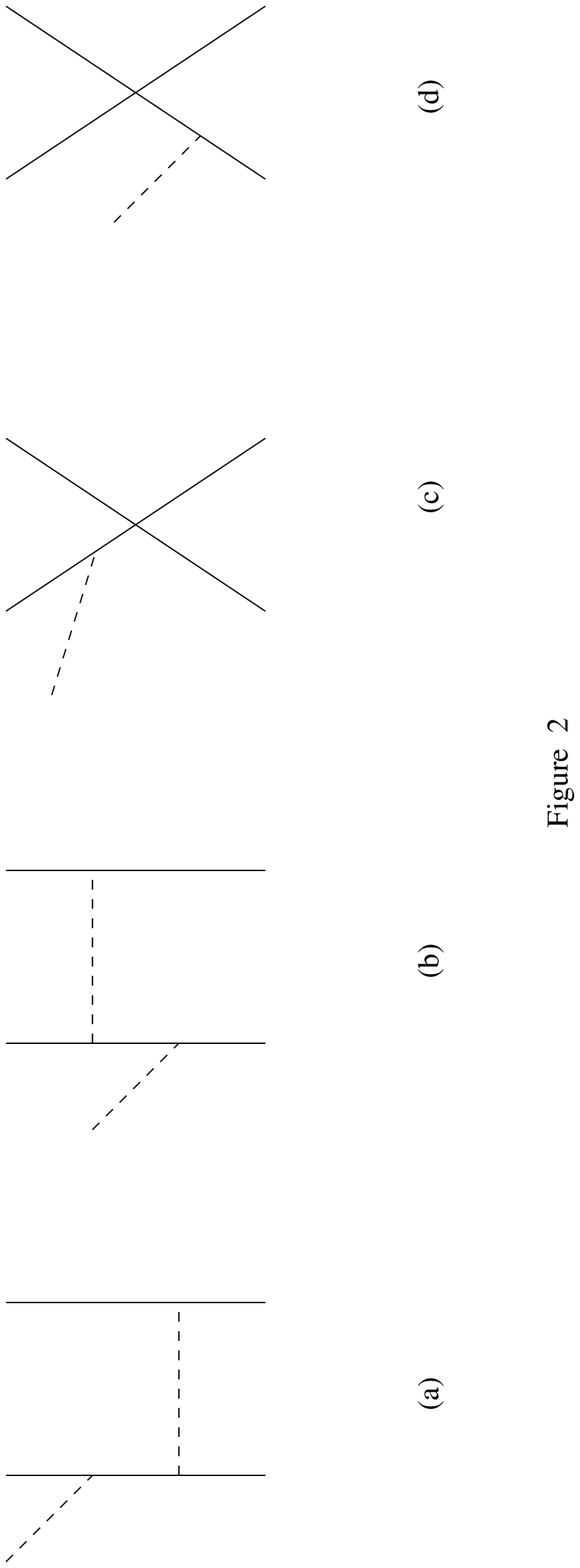}}
\end{figure}

\newpage

\begin{figure}
\centerline{\epsffile{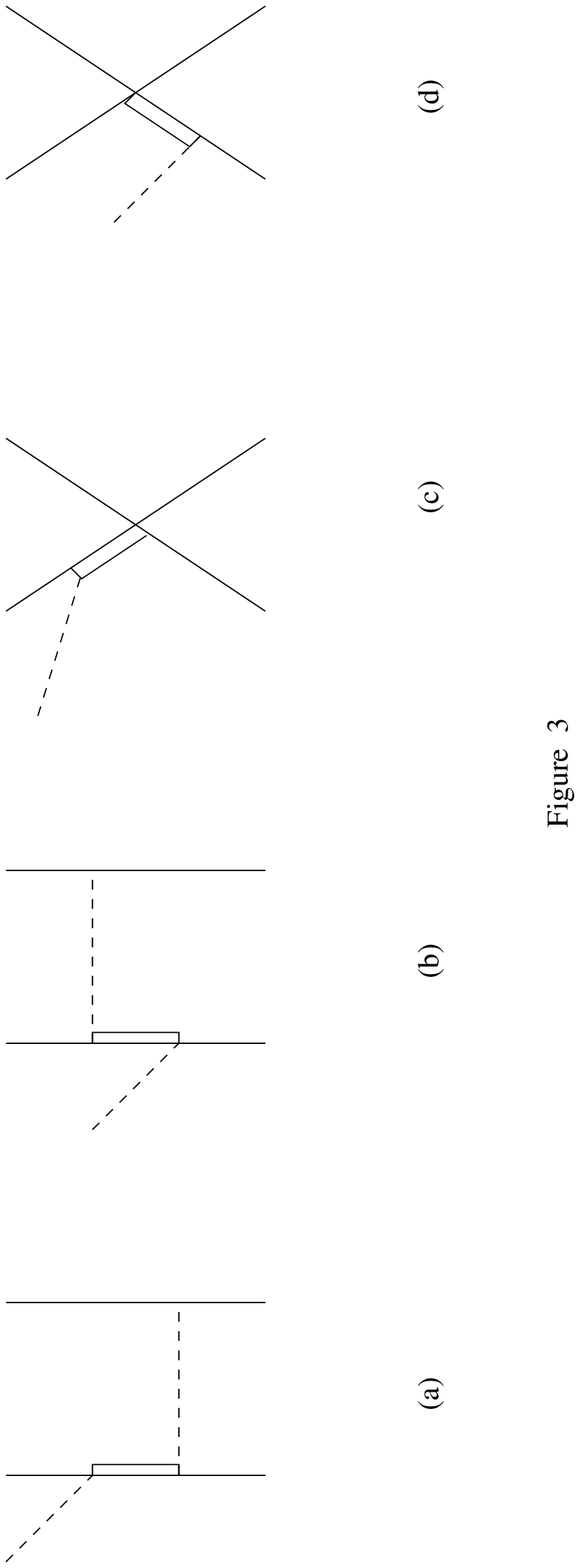}}
\end{figure}

\newpage

\begin{figure}
\centerline{\epsffile{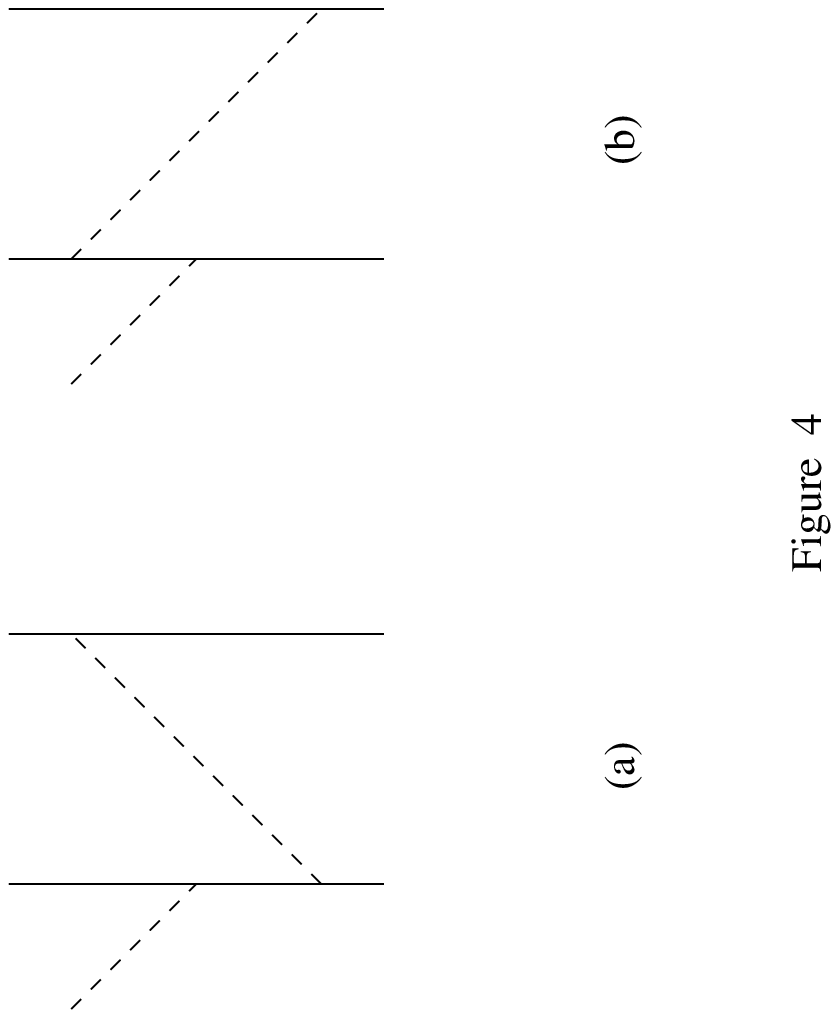}}
\end{figure}

\newpage

\begin{figure}
\centerline{\epsffile{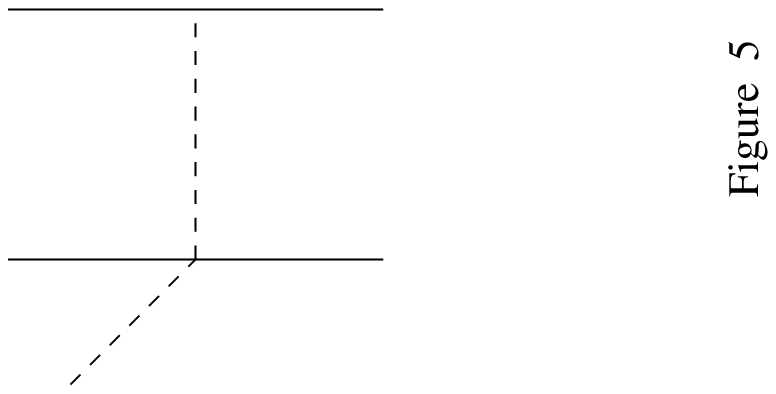}}
\end{figure}

\newpage

\begin{figure}
\centerline{\epsffile{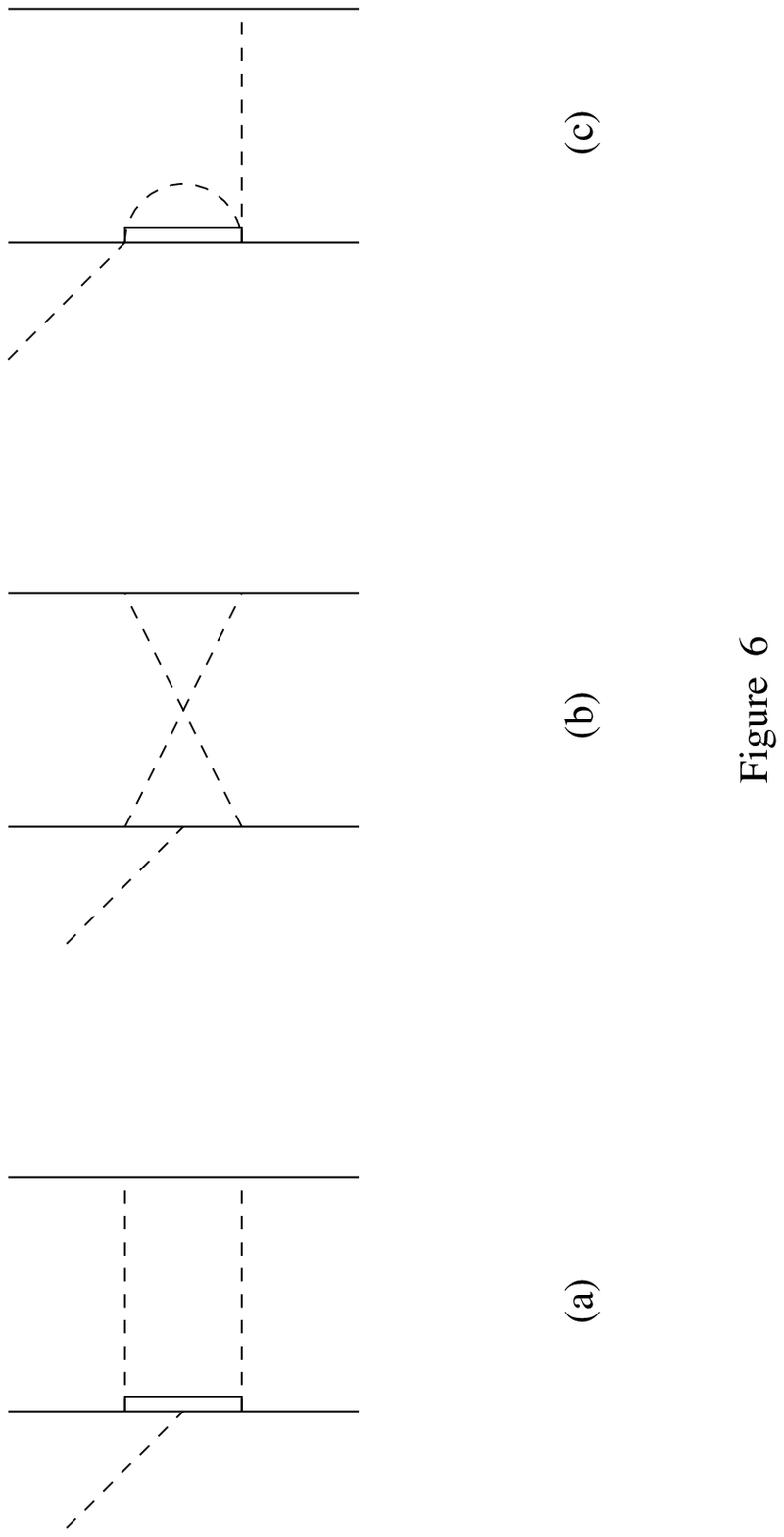}}
\end{figure}

\newpage

\begin{figure}
\centerline{\epsffile{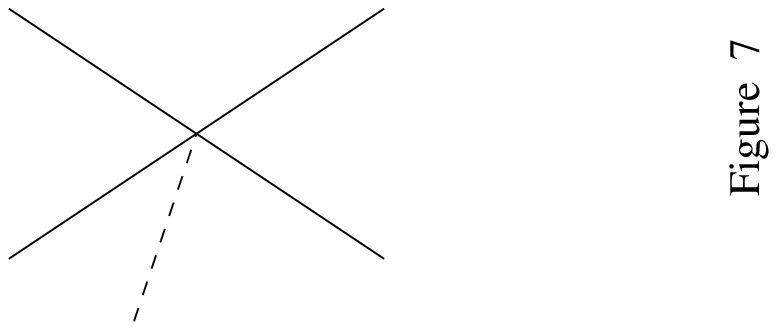}}
\end{figure}

\newpage

\begin{figure}
\centerline{\epsffile{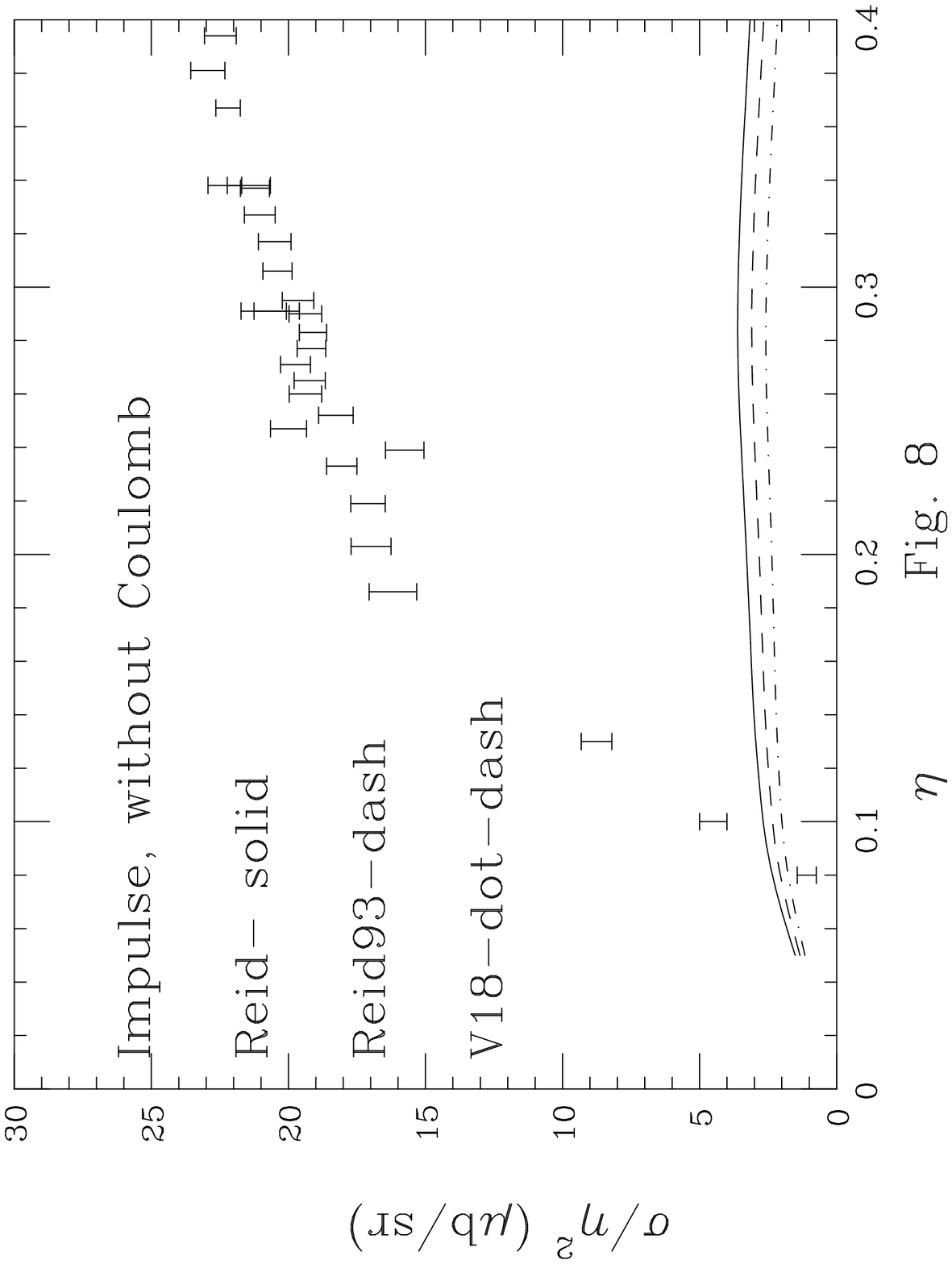}}
\end{figure}

\newpage

\begin{figure}
\centerline{\epsffile{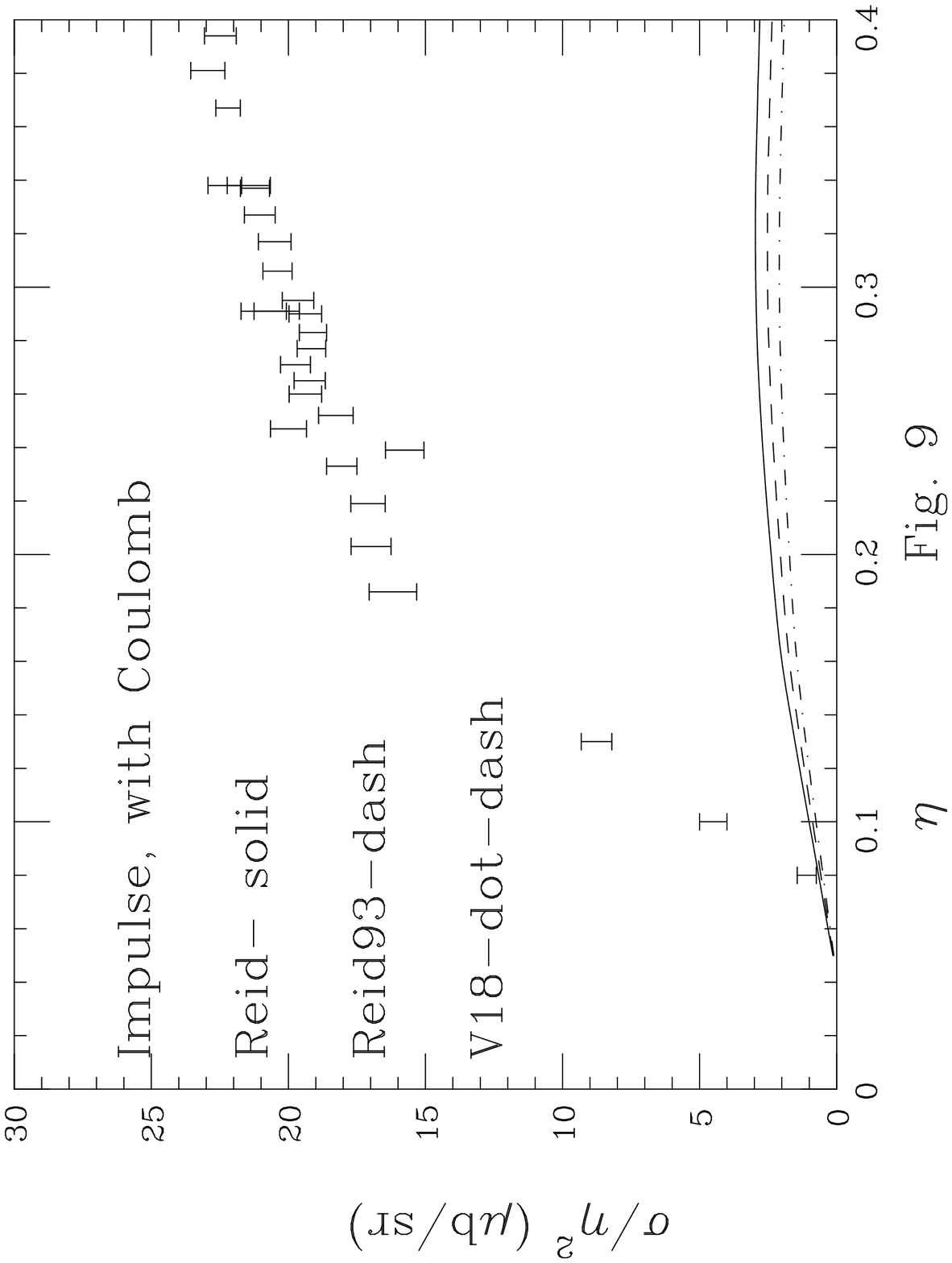}}
\end{figure}

\newpage

\begin{figure}
\centerline{\epsffile{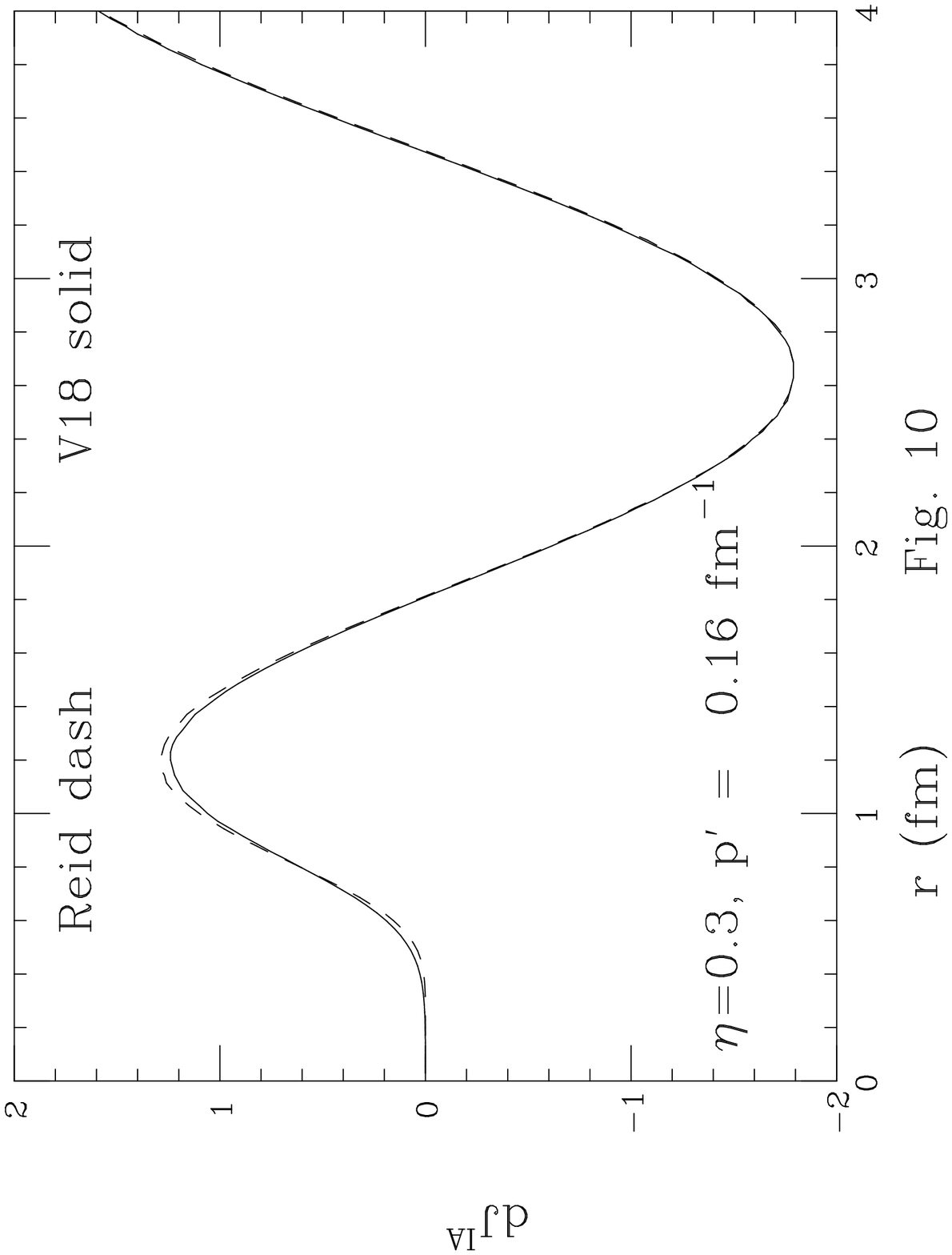}}
\end{figure}

\newpage

\begin{figure}
\centerline{\epsffile{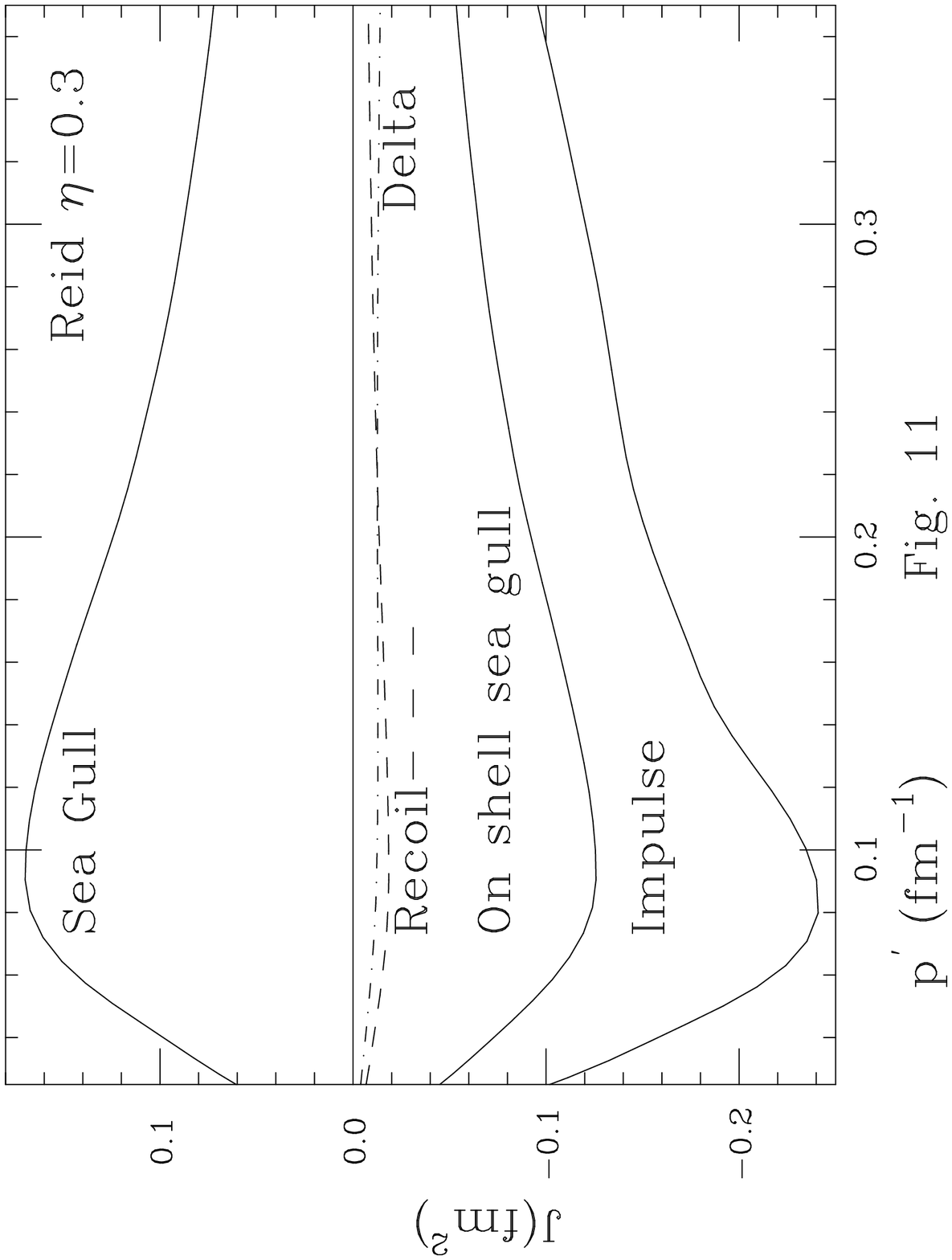}}
\end{figure}

\newpage

\begin{figure}
\centerline{\epsffile{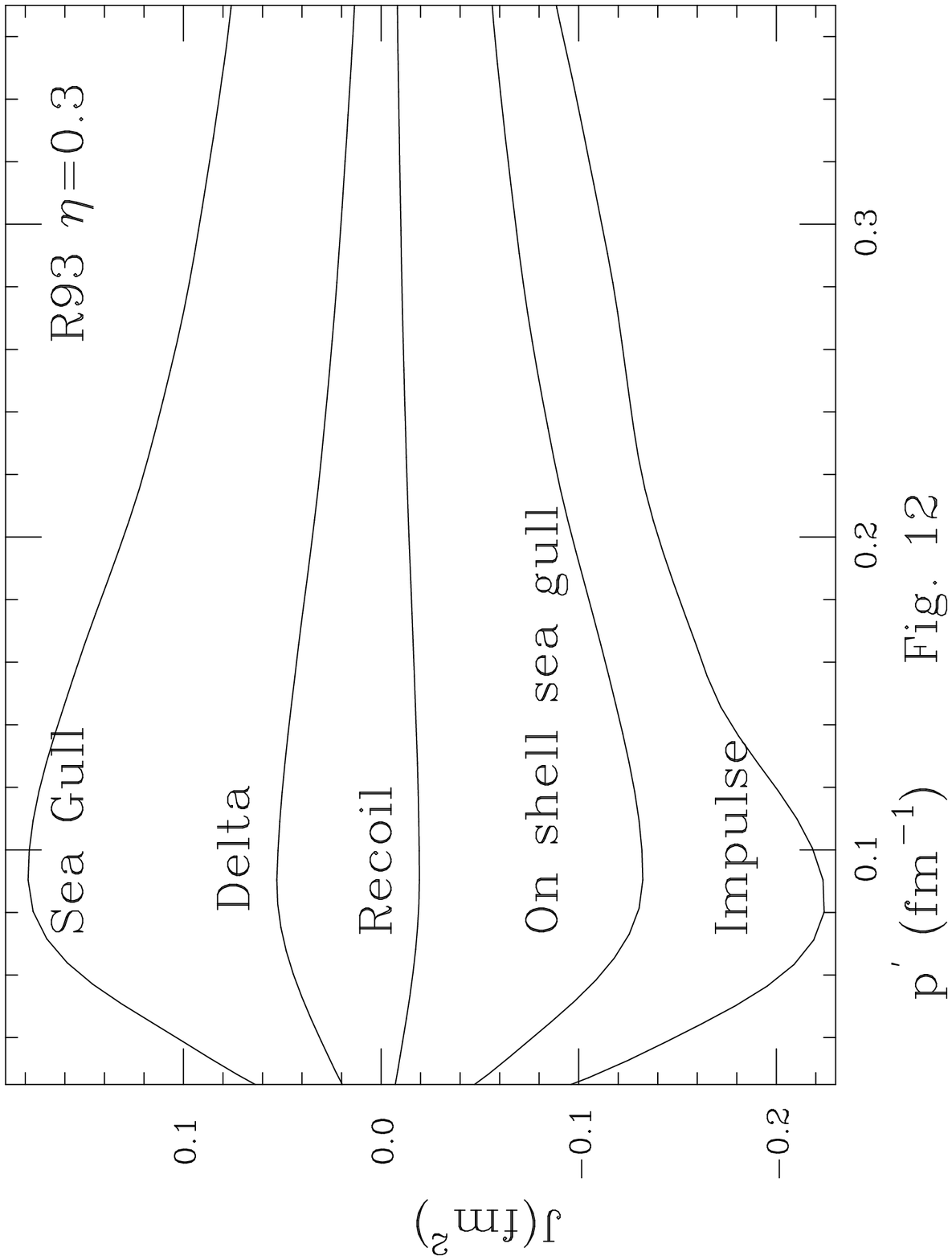}}
\end{figure}

\newpage

\begin{figure}
\centerline{\epsffile{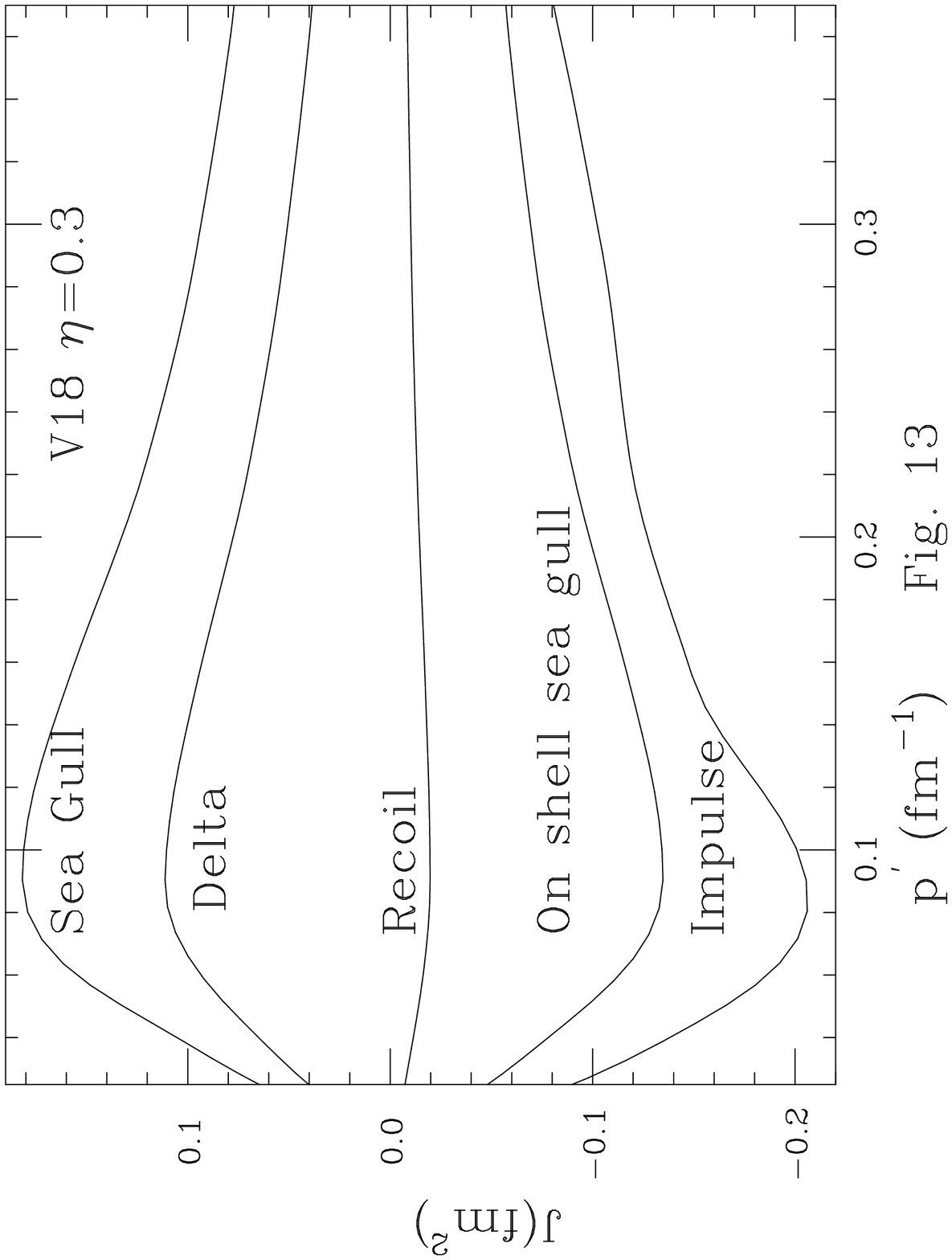}}
\end{figure}
                                               
\newpage

\begin{figure}
\centerline{\epsffile{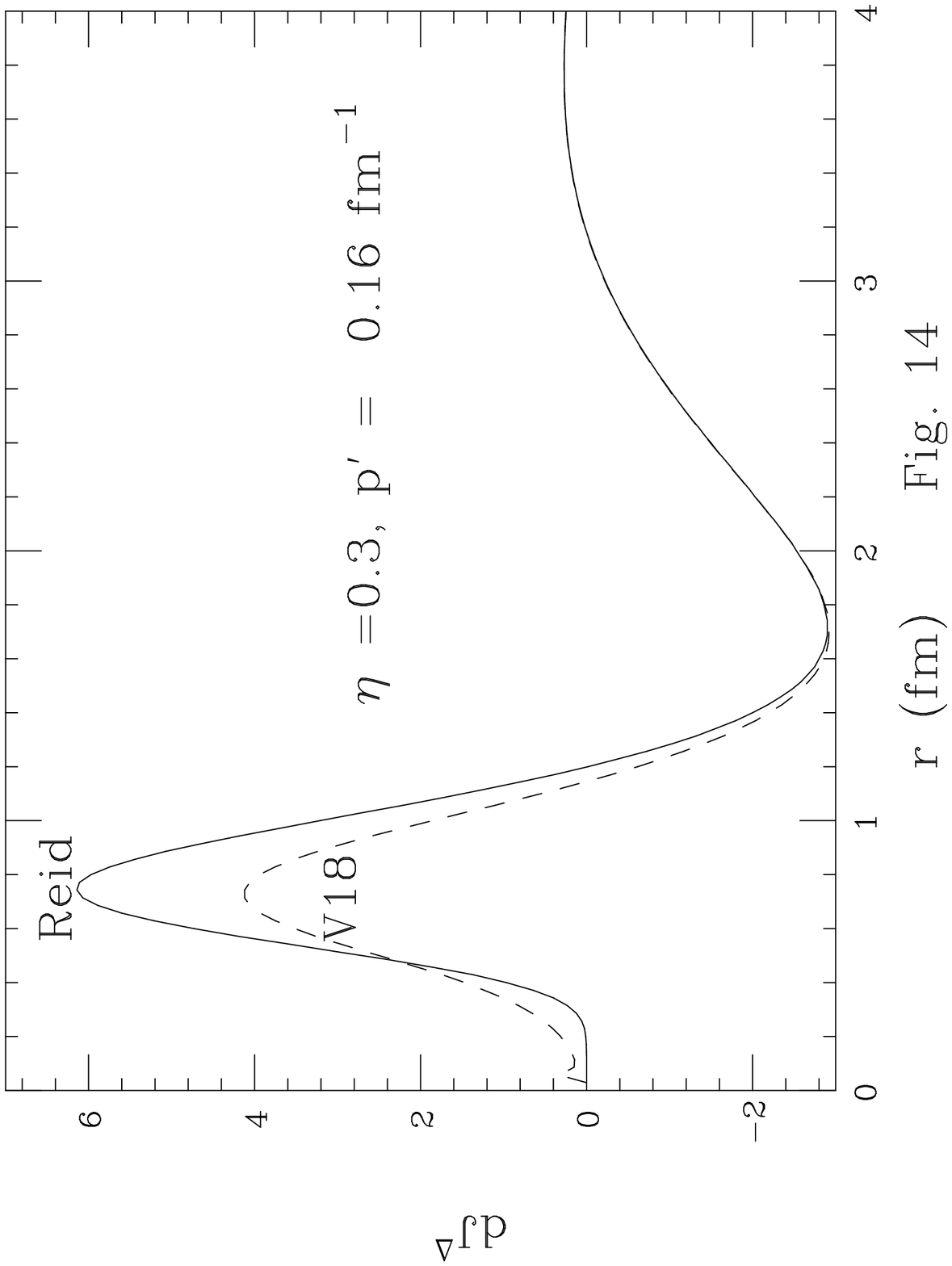}}
\end{figure}

\newpage

\begin{figure}
\centerline{\epsffile{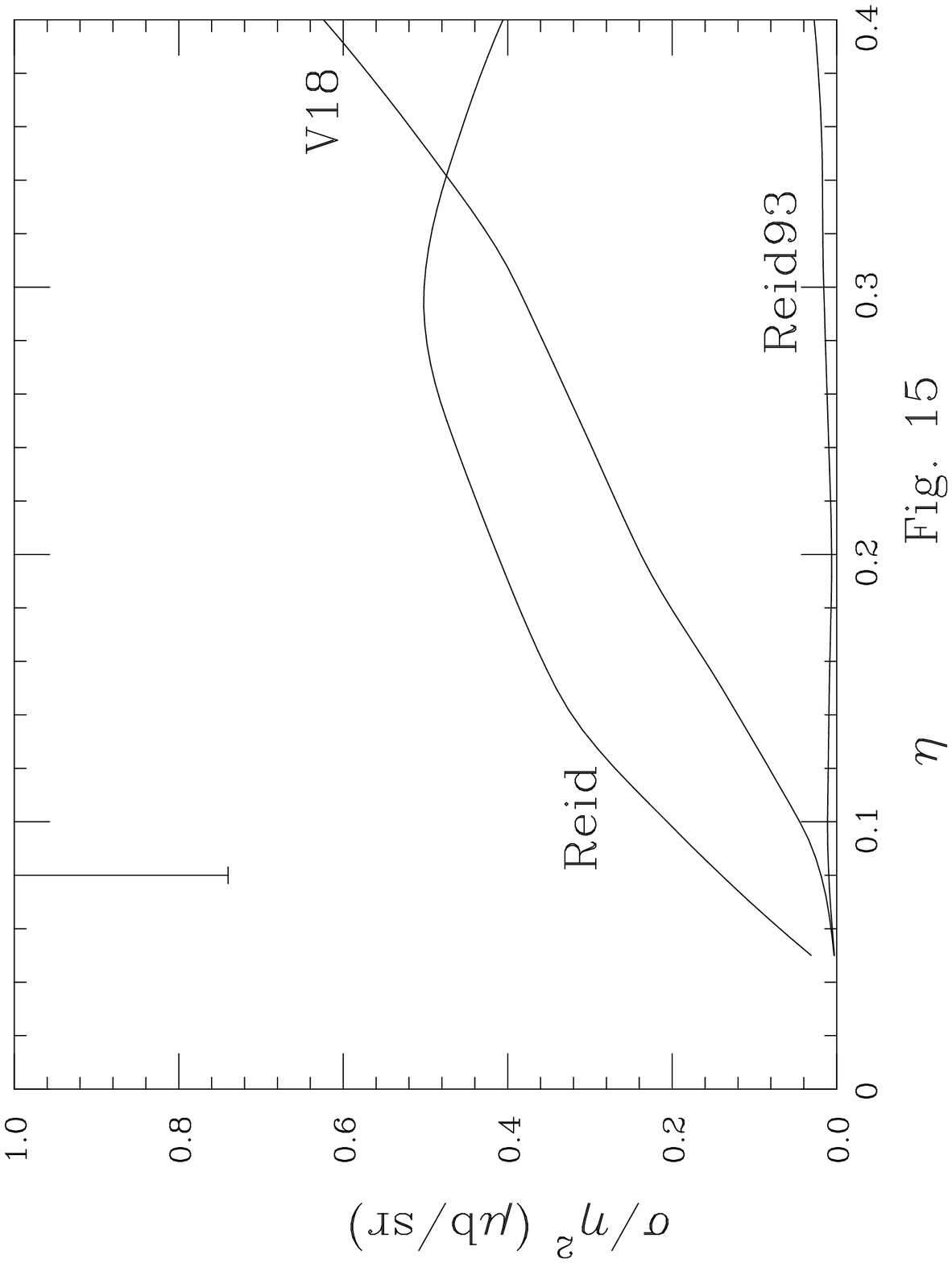}}
\end{figure}

\newpage

\begin{figure}
\centerline{\epsffile{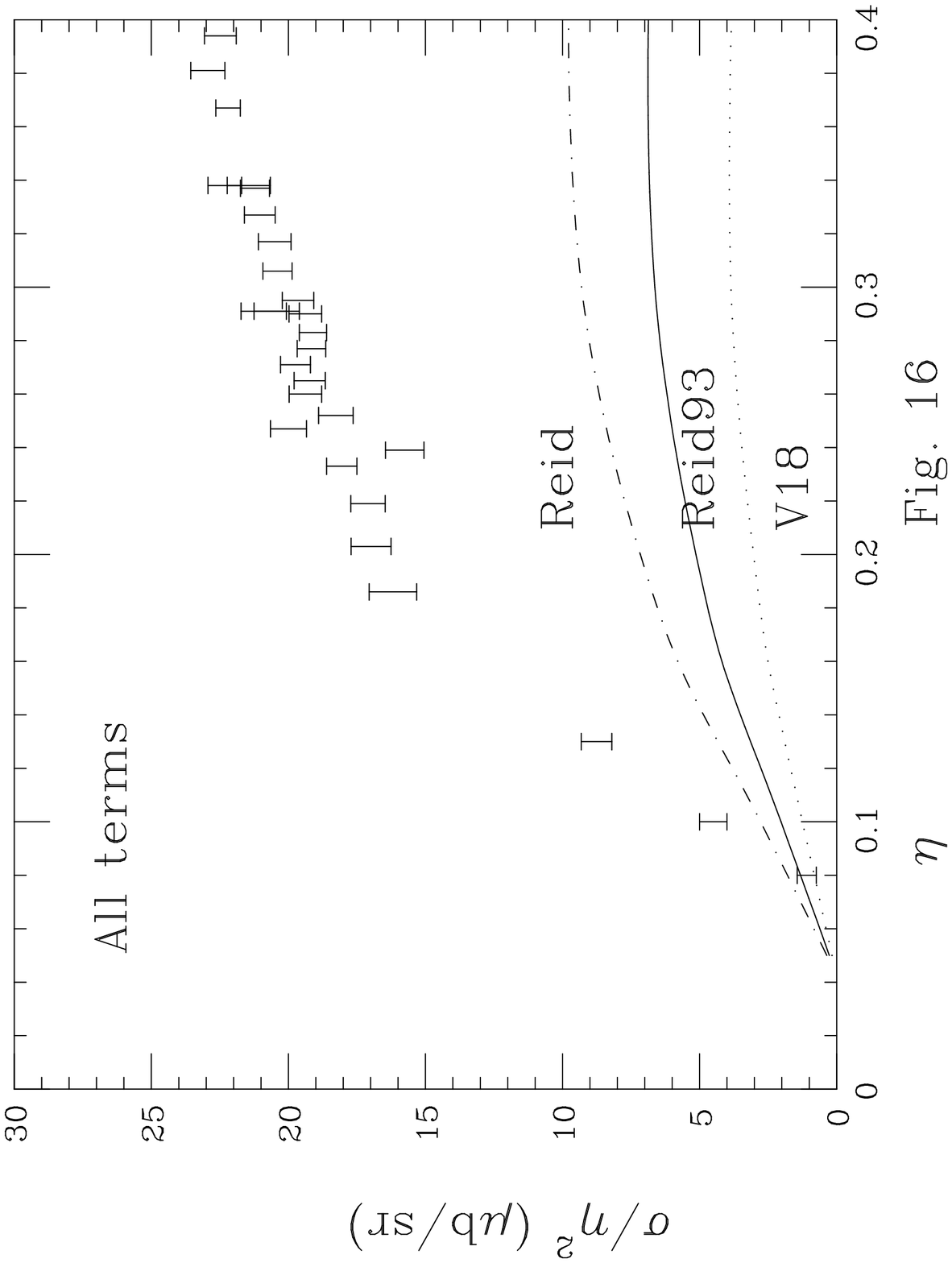}}
\end{figure}


\begin{references}
\bibitem{KR66} D.S. Koltun and A. Reitan, Phys. Rev. {\bf 141}, 
1413 (1966).

\bibitem{SSY69} M.E. Schillaci, R.R. Silbar, and J.E. Young,
Phys. Rev. {\bf 179}, 1539 (1969).

\bibitem{MS91} G.A. Miller and P.U. Sauer, Phys. Rev. C {\bf 
44}, R1725 (1991).

\bibitem{N92} J.A. Niskanen, Phys. Lett. {\bf B289}, 227 
(1992); Nucl. Phys. {\bf A298}, 417 (1978); Phys. Rev. C {\bf 43}, 36 (1991).


\bibitem{LR92} T.-S.H. Lee and D.O. Riska, 
Phys. Rev. Lett. {\bf 70}, 2237 (1992). 

\bibitem{HGM94} C.J. Horowitz, D.K. Griegel, and H.O. Meyer, Phys. 
Rev. C {\bf 49}, 1337 (1994).

\bibitem{HO95} E. Hern\'andez and E. Oset, Phys. Lett. {\bf B350}, 158 
(1995). 

\bibitem{julich} C. Hanhart, J. Haidenbauer, A. Reuber,
C. Sch\"{u}tz and J. Speth,  Phys. Lett. {\bf B358}, 21 (1995).

\bibitem{M90} H.O. Meyer {\it et al.}, Phys. Rev. Lett. {\bf 65},  
2846 (1990); Nucl. Phys. {\bf A539}, 633 (1992).

\bibitem{uppsala} A. Bondar {\it et al.}, Phys. Lett. {\bf B356}, 8 (1995).

\bibitem{W87} For a review of the Dirac phenomenology see 
S.J. Wallace, Ann. Rev. Nucl. Part. Sci. {\bf 37}, 267 (1987).

\bibitem{W79} S. Weinberg,  Physica {\bf 96A}, 327 (1979).

\bibitem{GL84} J. Gasser and H. Leutwyler, Ann. Phys. {\bf 158}, 142 (1984); 
Nucl. Phys. {\bf B250}, 465 (1985). 
 
\bibitem{JM91}  E. Jenkins and A.V. Manohar, Phys. Lett. {\bf B255}, 
558 (1991). 

\bibitem{BKM95} For a current review see
V. Bernard, N. Kaiser, and  U.-G. Meissner, Int. J. Mod. Phys. 
{\bf E4}, 193 (1995). 

\bibitem{W90} S. Weinberg, Phys. Lett. {\bf B251}, 288 (1990); 
Phys. Lett. {\bf B295}, 114 (1992). 

\bibitem{ORV94}  C. Ord\'o\~nez, L. Ray, and U. van Kolck,  
Phys. Rev. Lett. {\bf 72}, 1982 (1994); Univ. of Washington preprint
DOE/ER/40427-14-N95, hep-ph/9511380.

\bibitem{V94}  U. van Kolck, Phys. Rev. C {\bf 49}, 2932 (1994).

\bibitem{BLO95} 
U. van Kolck, G.A. Miller, J.L. Friar, and T.D. Cohen,
Bull. Am. Phys. Soc. {\bf 40}, 1629 (1995), presented 
at the DNP/APS Fall Meeting in Bloomington, Indiana, October 1995. 

\bibitem{SC95} B.-Y. Park, F. Myhrer, J.R. Morones, T. Meissner, and K. 
Kubodera, U. of South Carolina preprint USC(NT)-95-6, nucl-th/9512023.

\bibitem{CB92} T.D. Cohen and W. Broniowski, Phys. Lett. {\bf 
B292}, 5 (1992). 

\bibitem{C95a}  T.D. Cohen, Phys. Lett. {\bf 
B359}, 23 (1995).  

\bibitem{C95b} T.D. Cohen, U. of Maryland
preprint  96-057, hep-ph/9512275.

\bibitem{CWZ} S. Coleman, J. Wess, and B. Zumino, Phys. Rev. {\bf 
177}, 2239 (1969); C.G. Callan, S. Coleman, J. Wess, and B. Zumino,
{\it ibid.}, 2247.

\bibitem{machleidt} R. Machleidt, Adv. Nucl. Phys. {\bf 19}, 189 (1989). 

\bibitem{Reid} R.V.\ Reid, Ann.\ Phys.\ (NY) {\bf 50}, 411 (1968).

\bibitem{Reid93} J.L. Friar, G.L. Payne, V.G.J. Stoks, 
and J.J. de Swart, Phys.\ Lett. {\bf B311}, 4 (1993).

\bibitem{v18} R.B. Wiringa, V.G.J. Stoks, and R. Schiavilla,
Phys.\ Rev.\ C {\bf 51}, 38 (1995).

\bibitem{JN96} J.A. Niskanen, Phys. Rev. C {\bf 53}, 526 (1996).

\bibitem{pdb} Particle Data Group, Phys. Rev. D {\bf 50}, 1173 (1994).

\bibitem{weise} T. Ericson and W. Weise, {\it Pions and Nuclei}, 
                Clarendon Press, Oxford (1988).

\bibitem{jerry} G.A. Miller, B.M.K. Nefkens, and I. \v{S}laus,
Phys. Rep. {\bf 194}, 1 (1990); B.M.K. Nefkens, G.A. Miller and I. \v{S}laus,
Comm. Nucl. Part. Phys. {\bf 20}, 221 (1991).

\bibitem{M79} D.F. Measday and G.A. Miller,
Ann. Rev. Nucl. Part. Sci. {\bf 29}, 121 (1979).

\bibitem{F80} H.W. Fearing,
Prog. Part. Nucl. Phys. {\bf7}, 113 (1981).

\bibitem{jim} S.A. Coon and J.L. Friar, 
Phys.\ Rev.\ C {\bf 34}, 1060 (1986).

\end{references}
\end{document}